\pdfoutput=1

\documentclass[11pt]{article}

\usepackage[final]{acl}

\usepackage{times}
\usepackage{latexsym}

\usepackage[T1]{fontenc}

\usepackage[utf8]{inputenc}

\usepackage{microtype}

\usepackage{inconsolata}

\usepackage{graphicx}

\usepackage{amsmath} 
\usepackage{bbm}
\usepackage{array} 
\usepackage{booktabs}
\usepackage{multirow} 
\usepackage{xcolor}
\usepackage{adjustbox}
\usepackage{amssymb}
\usepackage{pifont}
\usepackage{algorithmic}
\usepackage{dsfont}
\newcommand{\cmark}{\ding{51}}
\newcommand{\xmark}{\ding{55}}
\definecolor{check}{rgb}{0.333, 0.755, 0.545}
\definecolor{mygray}{gray}{0.4}

\title{When Should Dense Retrievers Be Updated in Evolving Corpora? \\
Detecting Out-of-Distribution Corpora Using GradNormIR}

\author{First Author \\
  Affiliation / Address line 1 \\
  Affiliation / Address line 2 \\
  Affiliation / Address line 3 \\
  \texttt{email@domain} \\\And
  Second Author \\
  Affiliation / Address line 1 \\
  Affiliation / Address line 2 \\
  Affiliation / Address line 3 \\
  \texttt{email@domain} \\}

\author{
\textbf{Dayoon Ko}$^1$ \quad 
\textbf{Jinyoung Kim}$^1$ \quad 
\textbf{Sohyeon Kim}$^2$ \quad 
\textbf{Jinhyuk Kim}$^3$ \quad \\
\textbf{Jaehoon Lee}$^3$ \quad 
\textbf{Seonghak Song}$^3$ \quad 
\textbf{Minyoung Lee}$^3$ \quad 
\textbf{Gunhee Kim}$^1$ \\
\\
$^1$Seoul National University \quad $^2$POSTECH \quad $^3$Samsung SDS\\
\texttt{\footnotesize dayoon.ko@vision.snu.ac.kr \quad jiny1623@snu.ac.kr \quad sooohyun@postech.ac.kr \quad gunhee@snu.ac.kr}\\
\texttt{\footnotesize \href{https://github.com/dayoon-ko/gradnormir}{https://github.com/dayoon-ko/gradnormir}}
}

\begin{document}
\maketitle

\begin{abstract}
Dense retrievers encode texts into embeddings to efficiently retrieve relevant documents from large databases in response to user queries. However, real-world corpora continually evolve, leading to a shift from the original training distribution of the retriever. Without timely updates or retraining, indexing newly emerging documents can degrade retrieval performance for future queries. Thus, identifying when a dense retriever requires an update is critical for maintaining robust retrieval systems. In this paper, we propose a novel task of predicting whether a corpus is out-of-distribution (OOD) relative to a dense retriever before indexing. Addressing this task allows us to proactively manage retriever updates, preventing potential retrieval failures. We introduce \textbf{GradNormIR}, an unsupervised approach that leverages gradient norms to detect OOD corpora effectively. Experiments on the BEIR benchmark demonstrate that GradNormIR enables timely updates of dense retrievers in evolving document collections, significantly enhancing retrieval robustness and efficiency.
\end{abstract}

\section{Introduction}

With the exponential growth of digital content, information retrieval (IR) systems have become essential for delivering relevant information from massive document repositories \citep{bajaj2016ms, kwiatkowski2019natural}. Unlike traditional sparse retrieval methods \citep{robertson2009probabilistic, ramos2003using} that rely heavily on lexical overlap, dense retrievers \citep{karpukhin2020dense, izacardunsupervised} utilize semantic embeddings to better understand query intent and retrieve documents with similar conceptual meanings, thus overcoming the constraints of exact term matching. Consequently, dense retrievers have gained significant attention in tasks demanding high semantic precision, such as question answering and personalized search. During training, these dense retrievers are optimized to enhance embedding similarity between queries and relevant passages while reducing similarity for irrelevant ones \citep{karpukhin2020dense, izacardunsupervised}. Document embeddings are then precomputed and stored during indexing, enabling rapid retrieval during inference by identifying documents most similar to the test query embeddings.

\begin{figure}[t]
    \includegraphics[width=0.48\textwidth]{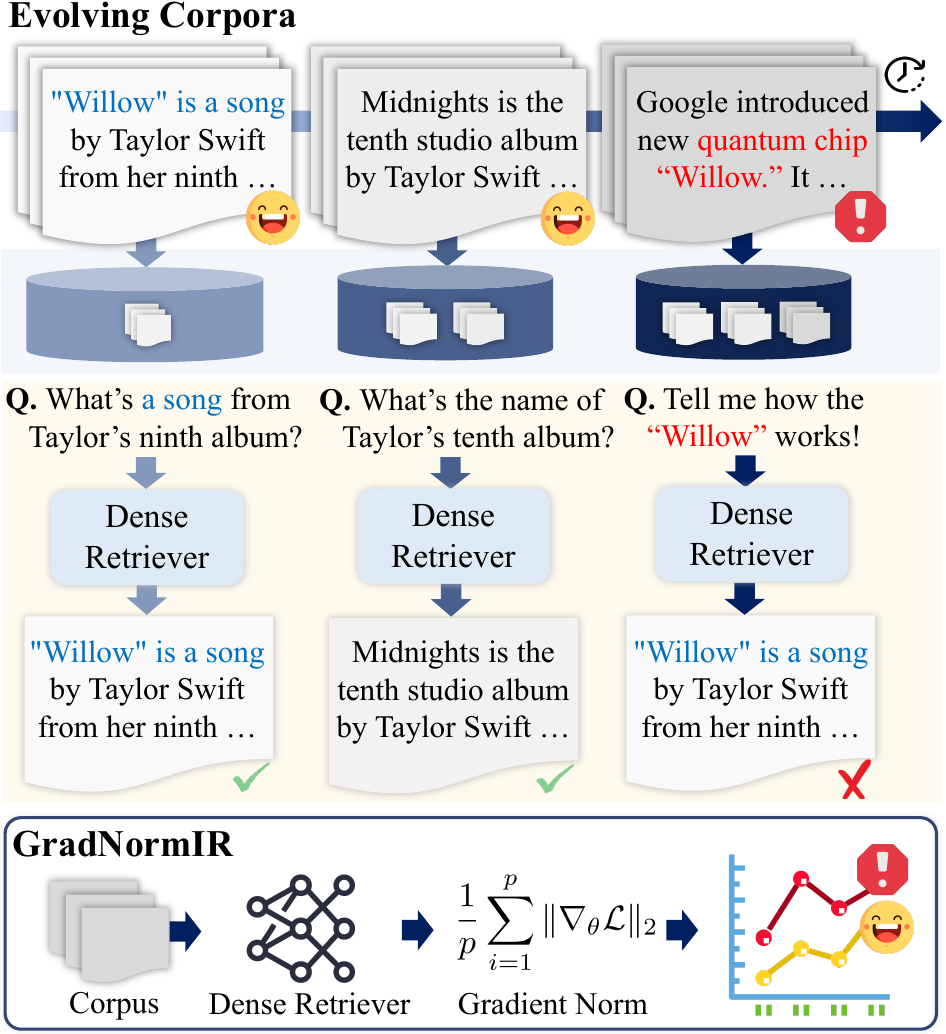}
    \caption{
        \textbf{Motivation}. In evolving corpora, indexing documents that dense retrievers fail to generalize to can severely degrade retrieval performance. Therefore, proactively detecting OOD corpora prior to indexing, without relying on available queries, is essential for maintaining retrieval effectiveness. To address this, we propose \textbf{GradNormIR}, an unsupervised method leveraging gradient norms to identify such OOD corpora.
    }
    \label{fig:intro}
    \vspace{-0.3cm}
\end{figure}

In the real world, corpora evolve rapidly due to technological advancements, societal changes, and emerging trends. This continuous evolution poses a substantial challenge for dense retrievers \citep{chen2023continual}, which often struggle to generalize effectively to unseen documents in zero-shot scenarios \citep{chen2022out}. The problem is particularly critical in retrieval-augmented generation (RAG) systems \citep{lewis2020retrieval}, where retriever performance directly impacts downstream tasks \citep{petronicontext, li2023large, ko-etal-2024-growover, kim-etal-2024-dynamicer}. For instance, consider the scenario illustrated in Figure. \ref{fig:intro}: when a new corpus about Google's quantum computing chip, \textit{\textcolor{red}{Willow}}, is introduced, a dense retriever trained on existing content, such as the song \textit{\textcolor{blue}{Willow}} by Taylor Swift, may erroneously retrieve irrelevant documents. A query like "Tell me how the \textit{\textcolor{red}{Willow}} works!" could mistakenly return information about the song \textit{\textcolor{blue}{Willow}} instead of the quantum chip. Anticipating when a retriever might fail due to such distributional shifts is crucial. Also, proactively identifying when to update the retriever ensures robustness and accuracy in dynamically evolving document streams.

This challenge closely relates to the OOD generalization problem. Several approaches in information retrieval (IR) aim to improve a retriever's performance on queries or documents significantly different from those encountered during training \citep{izacardunsupervised, wang2021gpl, chen2022out, yu2022coco, kasela2024desire, besta2024multi, chen2023continual}. One prominent approach utilizes a mixture-of-experts framework, employing a gating mechanism to select the most suitable expert retriever for each test query \citep{kasela2024desire,lee2024routerretriever}. However, these methods depend heavily on predefined expert retrievers trained offline with explicit domain knowledge and established domain boundaries, making them less flexible for dynamically evolving corpora. Determining the appropriate timing to introduce new experts for emerging content remains challenging.

To address this challenge, we introduce a novel practical task of \textit{predicting OOD corpora before indexing} for a given dense retriever. Identifying OOD corpora in advance signals when retriever updates are needed. By detecting such corpora before indexing, we can proactively select a more suitable dense retriever or promptly update the current one, safeguarding inference-time performance. To achieve this, we propose \textbf{GradNormIR}, an unsupervised method designed to detect OOD documents within a corpus without relying on queries. Inspired by successful applications in image classification, where gradient norms effectively detect OOD images and estimate test-time accuracy without labeled data \citep{huang2021importance, xie2024characterising}, GradNormIR leverages gradient norms from the contrastive loss to evaluate a retriever's generalizability on a given corpus. Specifically, we introduce novel sampling strategies to effectively assign positive and negative instances for computing the contrastive loss.

We assess our method on the BEIR benchmark, which encompasses multiple diverse datasets across various domains. Firstly, we demonstrate that GradNormIR effectively detects OOD documents likely to cause retrieval failures. Subsequently, we show that GradNormIR can select the most suitable retriever using only the corpus, without queries. Finally, we simulate evolving corpora using BEIR by sequentially introducing datasets following \citet{gelightweight}, and demonstrate how GradNormIR enables efficient retriever updates while maintaining performance. 
Our experiments validate both the importance of OOD detection for retrieval systems and GradNormIR's efficacy in adapting to evolving corpora.

In summary, our contributions are as follows:

\begin{enumerate}
    \item We introduce a novel task of predicting OOD corpora before indexing, facilitating efficient and effective retriever updates in evolving corpora.
    \item We propose GradNormIR, an unsupervised method that leverages gradient norms and novel sampling strategies to detect OOD documents and predict OOD corpora.
    \item On the BEIR benchmark, we demonstrate both the necessity of the proposed task for a robust retrieval system and the effectiveness of GradNormIR via three practical use cases.
\end{enumerate}

\section{Related Work}

\textbf{Information Retrieval.}
Recent advancements in text embeddings have significantly transformed the field of IR, particularly with the emergence of dense retrievers. The success of these models has primarily been driven by the availability of large-scale training datasets, including NQ \citep{kwiatkowski2019natural}, MS-MARCO \citep{bajaj2016ms}, HotpotQA \citep{yang2018hotpotqa}, and NLI \citep{gao2021simcse}. A notable example is DPR \citep{karpukhin2020dense}, which employs a dual-encoder architecture for open-domain question-answering, independently embedding queries and passages.

Unsupervised approaches have also garnered attention for their ability to enhance the generalization of dense retrievers. Contriever \citep{izacardunsupervised} enlarges pre-training data using unsupervised data augmentation techniques for contrastive learning. Similarly, E5 \citep{wang2022text} leverages weak supervision to create a large-scale dataset, CCPairs, filtered by consistency criteria. Recently, hybrid methods like BGE-M3 \citep{chen2024bge} have integrated dense, sparse, and multi-vector retrieval strategies via self-knowledge distillation, further advancing retrieval effectiveness.

\textbf{OOD Robustness.}
In IR, OOD robustness refers to a model's capacity to maintain retrieval effectiveness when exposed to documents that deviate significantly from its training distribution. A prominent benchmark to evaluate this robustness is BEIR \citep{thakur2beir}, which encompasses diverse retrieval tasks across multiple domains. Using BEIR, \citet{chen2022out} demonstrated that dense retrievers often underperform on OOD datasets compared to traditional lexical retrievers like BM25. In response, they proposed a hybrid model combining dense and sparse retrieval, achieving robust performance in zero-shot scenarios. Similarly, \citet{yu2022coco} showed that distribution shifts lead to a substantial decline in zero-shot accuracy of dense retrievers.

Various strategies have been proposed to improve OOD performance on unseen documents. Data augmentation has yielded promising results \citep{wang2021gpl, izacardunsupervised}. Architectural adaptations have also been explored; mixture-of-experts frameworks \citep{kasela2024desire, lee2024routerretriever} and multi-head RAG models \citep{besta2024multi} adjust retrieval strategies based on domain characteristics. Moreover, \citet{khramtsova2023selecting} investigated zero-shot retrieval methods for selecting the most suitable retriever, while \citet{khramtsova2024leveraging} proposed leveraging LLM-generated pseudo-queries to rank dense retrievers. Other studies \citep{cai2023l2r, chen2023continual} have adopted continual learning methods to handle dynamic corpora without forgetting previously acquired knowledge. For instance, memory-based approaches \citep{cai2023l2r} maintain backward compatibility with existing document embeddings, and incremental indexing strategies \citep{chen2023continual} dynamically update document indices in generative retrievers to accommodate new information.

Despite these advancements, few studies have explicitly examined how to identify OOD documents from the perspective of dense retrievers. Layer-wise score aggregation methods, such as those proposed by \citet{darrin2024unsupervised}, combine anomaly scores from each encoder layer to provide more accurate detection. However, this approach is primarily tailored to text classification tasks, whereas our work explicitly addresses evaluating the generalizability of dense retriever models across evolving document corpora.

\section{Problem Statement}
\label{sec:ood-definition}

\subsection{OOD Robustness in IR}

OOD robustness refers to the ability of a model to maintain effective performance when encountering data distributions different from the training distribution. Following \citet{liu2024robust}, we define OOD robustness in IR formally as:

\begin{align}
|\mathcal{R}_M(f_\theta; \mathcal{D}_\text{test}, K) - \mathcal{R}_M(f_\theta; \tilde{\mathcal{D}}_\text{test}, K)| \leq \delta, \notag \\
\text{where } \mathcal{D}_\text{train}, \mathcal{D}_\text{test} \sim \mathcal{G},\;\tilde{\mathcal{D}}_\text{test} \sim \tilde{\mathcal{G}}.
\label{eq:ood-robust}
\end{align}

Here, $f_\theta$ is a dense retriever trained on the training set $\mathcal{D}_\text{train}$ drawn from the original distribution $\mathcal{G}$. $\mathcal{R}_M(f_\theta; \mathcal{D}, K)$ denotes the retrieval performance metric (e.g., Recall@K) for the top-$K$ retrieved results by $f_\theta$, and $\delta$ represents an acceptable error margin. The test set $\mathcal{D}_\text{test}$ is drawn from the original distribution $\mathcal{G}$, while the new test set $\tilde{\mathcal{D}}_\text{test}$ originates from a distinct distribution $\tilde{\mathcal{G}}$. A retriever $f_\theta$ satisfying Eq.~(\ref{eq:ood-robust}) is considered $\delta$-robust against OOD data under metric $M$.

\subsection{OOD Document}

Based on the OOD robustness, we first define an ideal criterion for identifying an OOD document, assuming the availability of queries for evaluation. Given a relevant query-document pair $(q,d)$ with $d \in \mathcal{C}$, a document $d$ is considered OOD if the retriever $f_\theta$ fails to retrieve $d$ within the top-$K$ results:
\begin{equation}
\text{OOD}(d; f_\theta, q, \mathcal{C}) = \mathds{1}[d \notin f_\theta(q, \mathcal{C})],
\end{equation}

\noindent where $\mathds{1}$ is an indicator function. This definition directly reflects the retriever's inability to generalize to the document $d$. 

In practice, however, queries are often unavailable when a corpus is introduced. Thus, our proposed method $\mathcal{M}$ (see Section~\ref{sec:approach}), named \textbf{GradNormIR}, predicts such OOD documents proactively in an \textit{unsupervised} manner, without relying on labeled query-document pairs. The labeled query-document pairs are used only for evaluating the effectiveness of our unsupervised predictions.

\subsection{OOD Corpus}

Based on the predicted OOD documents, we aim to determine whether a given corpus $\mathcal{C}$ itself is OOD for the retriever $f_\theta$. The likelihood of $\mathcal{C}$ being OOD can be quantified by the proportion of predicted OOD documents it contains:
\begin{equation}
r(\mathcal{C}) = \frac{|\tilde{\mathcal{C}}|}{|\mathcal{C}|},\ \ \text{with}\ \ \tilde{\mathcal{C}} = \{ d \in \mathcal{C} \mid \mathcal{M}(d; f_\theta, \mathcal{C}) = 1 \}.
\label{eq:ood-corpus}
\end{equation}

A corpus $\mathcal{C}$ is classified as OOD if the ratio $r(\mathcal{C})$ exceeds a predefined threshold $\gamma$. This criterion enables proactive detection of corpus-level distributional shifts, facilitating timely retriever updates to maintain retrieval effectiveness.

\vspace{0.7cm}
\section{Approach}
\label{sec:approach}
We first review prior work leveraging gradient norms in image classification. We then introduce GradNormIR, an unsupervised method to proactively detect OOD documents and consequently identify OOD corpora.

\subsection{Preliminary of Gradient Norm} 
Previous studies utilize the gradient norm as an indicator of model performance and uncertainty in image classification. GradNorm \citep{huang2021importance} estimates uncertainty by computing the gradient norm derived from the KL divergence between the softmax output and a uniform distribution, identifying smaller gradient norms with higher uncertainty (OOD images). GDScore \citep{xie2024characterising} estimates test-time accuracy in an unsupervised manner by pseudo-labeling the input, computing cross-entropy loss, and measuring the gradient vector norm of the final layer.

Unlike these methods, our work applies gradient norms in information retrieval (IR) to proactively detect OOD documents by employing novel sampling strategies to compute the gradient norm of contrastive loss.

\subsection{GradNormIR}\label{sec:gradnormir}

We use the gradient norm to detect OOD documents in a corpus. To get the gradient norm, we need to calculate the loss. Dense retrievers are usually trained with InfoNCE loss, defined as
\begin{align}
     &\mathcal{L}_{\text{InfoNCE}} = - \log \frac{e^{s(q, d^+)/\tau}}{e^{s(q, d^+)/\tau} + \sum_{i=1}^{N} e^{s(q, d_i^-)/\tau}}, \nonumber
\end{align}
where $s(q, d) = \cos\left(f_\theta(q), f_\theta(d)\right)$ is the cosine similarity between query $q$ and document $d$. $f_\theta(\cdot)$ is  the last hidden layer’s output, and $\tau$ is a temperature parameter. 

When a new corpus $\mathcal{C}$ is given, user queries are not yet available, making it challenging to compute the gradients. Therefore, we consider each document $d$ as a query and assign pseudo-labels of positives and negatives to other documents $\mathcal{C}\setminus\{d\}$ that are relevant or irrelevant with $d$, respectively. Instead of using external trained models for such labeling, we obtain pseudo-labels directly from the retriever's own internal similarity scores. 

\textbf{Query Representation with Dropout.} As discussed, every $d$ is regarded as a document query. To better reflect the retriever's generalizability in the gradient norm, we introduce perturbations to the representation of $d$. Following \citet{jeong2022augmenting}, we apply stochastic dropout to randomly mask some parts of $d$'s representation $f_\theta(d)$. 
If $f_\theta(d)$ generalizes well to $d$, masking some tokens in its embedding has little impact on selecting its positive and negative samples. Otherwise, if the retriever poorly generalizes to document $d$, such masking induces significant shifts in the embedding space, causing incorrect positive and negative sample selections, and thereby yielding larger gradient norms.

Specifically, we first encode the document query with the last hidden state $h = f_\theta(d)$. We then randomly mask the hidden state; the mask $m$ is sampled from a Bernoulli distribution:
\begin{align}
    h' = h \odot m, \text{ where } m \sim \text{Bernoulli}(p), \nonumber
\end{align}
where $\odot$ denotes element-wise multiplication. Finally, we obtain the perturbed document query $d'$ by applying pooling on $h'$. 

\begin{figure}[t]
    \includegraphics[width=0.49\textwidth]{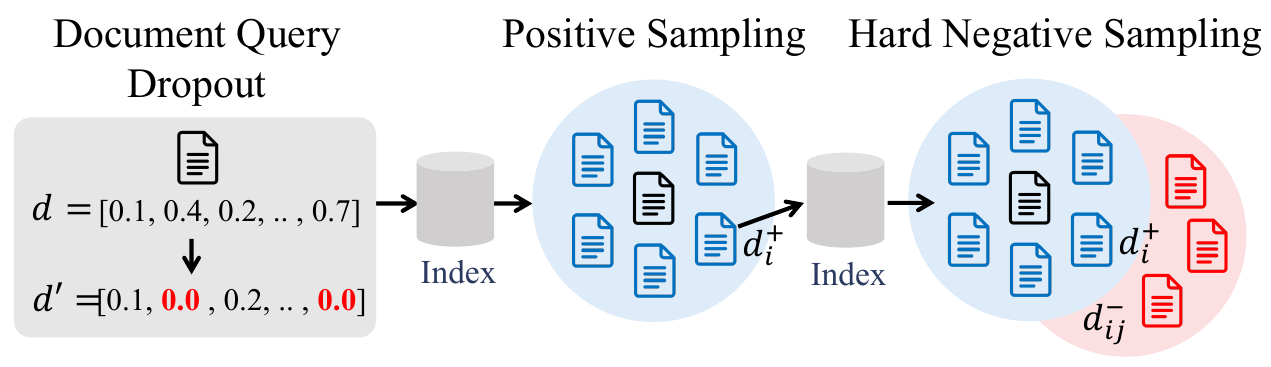}
    \caption{
        Dropout for the document query representation along with positive and hard negative sampling. 
    }
    \label{fig:approach}
    \vspace{-0.3cm}
\end{figure}

\textbf{Positive and Negative Sampling.} To effectively compute the gradient norm without actual queries, we introduce an internal pseudo-labeling approach inspired by \citep{xie2024characterising}. The overall process is illustrated in Figure~\ref{fig:approach}. Specifically, given a perturbed document query representation $d'$, we first retrieve the top-$k$ nearest documents from $\mathcal{C}\setminus\{d\}$ using $k$-nearest neighbors ($k$-NN). These documents form the positive candidate pool $D^+(d)$, assuming that documents with higher similarity scores are likely to be relevant.

From this positive pool, we further select the top-$p$ most similar documents $\{ d_1^+, \dots, d_p^+ \}$ as positives for computing the loss, ensuring precise estimation of the retriever's generalizability. The negative pool $D^-(d)$ comprises the remaining documents, assumed to be irrelevant or less relevant to $d$.

For negative samples, previous work \citep{zhan2021optimizing} has demonstrated that using hard negatives—documents similar to positives but irrelevant to the query—improves model sensitivity and performance. Therefore, we adopt a hard negative sampling strategy: for each positive $d_i^+$, we select its top-$n$ nearest documents $\{d_{i1}^-, \dots, d_{in}^-\}$ from the negative pool $D^-(d)$. These hard negatives increase gradient sensitivity, enabling effective detection of documents for which the retriever generalizes poorly.

\textbf{Gradient Norm.} We compute the gradient norm of $d$ based on the InfoNCE loss gradients with respect to the retriever parameters $\theta$:
\begin{align}
    \nabla\mathcal{L}_{\theta} = - \nabla_{\theta}\log \frac{e^{{s(d, d_i^+)}/\tau}}{e^{{s(d, d_i^+)}/\tau} + \sum_{j=1}^ne^{{s(d, d_{ij}^-)}/\tau}}, \nonumber
\end{align}
where $d_i^+$ is a positive sample and $d_{ij}^- \in D^-(d_i^+)$ are the corresponding hard negatives. The gradient norm measures the sensitivity of the retriever parameters to these samples.

Finally, the average gradient norm across all positive samples \(\{d_i^+\}_{i=1}^p\) is
\begin{equation}\label{eq:gradnormir}
    \text{GradNormIR} = \frac{1}{p} \sum_{i=1}^p\|\nabla_{\theta}\mathcal{L}\|_2,
\end{equation}

\noindent
where \( \|\mathcal{\cdot}\|_2 \) denotes the \( \mathcal{L}_2 \)-norm. This average gradient norm serves as a measure of the retriever's generalizability on $d \in \mathcal{C}$. A higher value indicates a greater sensitivity and potentially less stability when adapting to $d$. 

\textbf{Predicting OOD Documents.} 
We classify document $d$ as OOD if its gradient norm exceeds a threshold derived from the median gradient norm of known in-domain documents. The median provides robustness against outliers, ensuring reliable and stable OOD detection \citep{leys2013detecting}. We describe threshold selection as well as other hyperparameters in Sec~\ref{sec:exp-setup}.

\section{Experiments}
\vspace{-0.1cm}

\begin{table*}[t!]
\begin{center}
\renewcommand{\arraystretch}{0.9}
\resizebox{\linewidth}{!}{
\begin{tabular}{p{1.7cm} >{\arraybackslash}p{3cm} > {\centering\arraybackslash}p{1.3cm} >{\centering\arraybackslash}p{1.7cm} >{\centering\arraybackslash}p{1.3cm} >{\centering\arraybackslash}p{1.3cm} >{\centering\arraybackslash}p{1.3cm} >{\centering\arraybackslash}p{1.3cm} >{\centering\arraybackslash}p{1.3cm} >{\centering\arraybackslash}p{1.3cm} >{\centering\arraybackslash}p{1.3cm} >{\centering\arraybackslash}p{1.3cm}|>{\centering\arraybackslash}p{1.3cm}} 
    \toprule
    \textbf{Retriever} & \textbf{Documents} & ArguAna & C-FEVER & DBPedia & FiQA & NFCorpus & Quora & Scidocs & SciFact & COVID & Touch\'e & Avg ($\downarrow$) \\ \midrule
    \multirow{5}{*}{\textbf{BGE}} 
        & All & 99.68 & 79.96 & 59.67 & 80.25 & 21.39 & 99.68 & 72.33 & 99.76 & 16.53 & 98.45 & 73.48 \\ \cmidrule{2-13} 
        & OOD w/ Layerwise & \textbf{99.01} & 86.14 & 45.48 & 79.73 & 22.35 & 99.78 & 61.95 & 100.0 & 15.34 & 93.33 & 70.31  \\
        & OOD w/ IPQ & 100.0 & 74.65 & 55.02 & 81.75 & 19.11 & 100.0 & 82.24 & \textbf{99.72} & 15.60 & 100.0 & 72.81 \\
        & OOD w/ GenQuery & 100.0 & 86.87 & 75.48 & 79.70 & 21.42 & \textbf{98.96} & 62.13 & 100.0 & 15.97 & 97.40 & 73.79 \\
        & OOD w/ Ours & \textbf{99.01} & \textbf{61.14} & \textbf{31.49} & \textbf{79.16} & \textbf{18.36} & 99.71 & \textbf{56.97} & 100.0 & \textbf{15.22} & \textbf{89.19} & \textbf{65.03} \\ \midrule
    \multirow{5}{*}{\textbf{Contriever}} 
        & All & 96.79 & 72.40 & 56.76 & 59.83 & 18.66 & 98.83 & 55.26 & 98.25 & 9.14 & 96.14 & 66.06 \\ \cmidrule{2-13} 
        & OOD w/ Layerwise & 93.83 & 68.67 & 49.14 & 56.61 & 17.85 & 99.10 & 51.75 & 98.36 & 8.26 & 95.34 & 63.89 \\
        & OOD w/ IPQ & 93.83 & 69.11 & 48.37 & 57.72 & 18.11 & 99.17 & 51.39 & 98.70 & 8.25 & 94.04 & 63.87 \\
        & OOD w/ GenQuery & \textbf{90.12} & 72.23 & 65.87 & \textbf{54.99} & 18.53 & \textbf{97.28} & 51.17 & 98.65 & \textbf{7.45} & 93.33 & 64.96 \\
        & OOD w/ Ours & 91.36 & \textbf{63.92} & \textbf{40.63} & 56.12 & \textbf{17.11} & 98.75 & \textbf{50.64} & \textbf{97.78} & 8.09 & \textbf{90.17} & \textbf{61.46} \\ \midrule
    \multirow{5}{*}{\textbf{E5}} 
        & All & 99.68 & 76.42 & 55.56 & 74.85 & 18.03 & 99.67 & 61.49 & 98.49 & 15.81 & 97.75 & 70.00\\ \cmidrule{2-13} 
        & OOD w/ Layerwise & 100.0 & 75.46 & 47.01 & \textbf{74.38} & 19.04 & 99.65 & 55.53 & \textbf{98.43} & 15.96 & 97.81 & 68.33 \\
        & OOD w/ IPQ & \textbf{98.91} & 75.96 & 49.54 & 74.61 & 18.01 & 99.83 & 58.57 & 98.45 & \textbf{15.68} & 97.24 & 68.68 \\ 
        & OOD w/ GenQuery & \textbf{98.91} & 80.15 & 69.33 & 74.41 & 18.54 & \textbf{99.51} & 58.05 & 98.53 & 15.79 & 98.03 & 71.13 \\
        & OOD w/ Ours & 99.45 & \textbf{69.19} & \textbf{29.74} & 74.47 & \textbf{17.02} & 99.68 & \textbf{55.50} & 98.56 & 15.85 & \textbf{96.43} & \textbf{65.59} \\ \midrule
    \multirow{5}{*}{\textbf{GTE}} 
       & All & 99.68 & 80.37 & 60.85 & 75.76 & 22.48 & 99.57 & 72.66 & 99.52 & 17.53 & 99.25 & 73.55\\ \cmidrule{2-13} 
       & OOD w/ Layerwise & 100.0 & 82.81 & 56.66 & 76.17 & 21.48 & 99.87 & 66.14 & \textbf{99.47} & 14.98 & 99.82 & 71.74 \\
       & OOD w/ IPQ & 100.0 & 84.59 & 66.26 & 75.27 & 19.82 & 99.83 & 68.45 & 99.50 & \textbf{14.81} & 99.82 & 72.84  \\
       & OOD w/ GenQuery & 100.0 & 84.20 & 76.80 & \textbf{70.58} & 21.33 & \textbf{98.71} & 65.95 & 99.73 & 16.16 & 99.63 & 73.31 \\
       & OOD w/ Ours & \textbf{93.75} & \textbf{70.83} & \textbf{51.24} & 71.02 & \textbf{19.22} & 99.60 & \textbf{65.22} & 100.0 & 16.56 & \textbf{98.72} & \textbf{68.62} \\
    \bottomrule
\end{tabular}
}
\caption{
    Comparison of OOD document detection across different retriever models on the BEIR benchmark. A lower Document Retrieval Rate value, defined in Eq.(\ref{eq:acc}), indicates more accurate OOD detection.
}
\vspace{-0.6cm}
\label{tab:exp-prediction}
\end{center}
\end{table*}

We conduct three sets of experiments to evaluate our approach. First, Next, we verify that GradNormIR's OOD detection is effective for selecting the most suitable retriever. Next, we make sure that GradNormIR's OOD detection is useful in selecting the most suitable retriever, even without any queries. Finally, in evolving corpora, we demonstrate that GradNormIR enables efficient continuous retriever updates by selectively retraining it only on the predicted OOD corpus. We also present an ablation study for several hyperparameters in Appendix \ref{app:ablation}.

\subsection{Experimental Setup}\label{sec:exp-setup}

\textbf{Dense Retrievers.}\label{sec:exp-prediction-retrievers}
We evaluate several state-of-the-art dense retriever models, including BGE \citep{bge_embedding}, Contriever \citep{izacardunsupervised}, E5 \citep{wang2024multilingual}, and GTE \citep{li2023towards}. 

\textbf{Dataset.}
The BEIR benchmark \citep{thakur2beir} provides a diverse collection of datasets for evaluating retriever models across multiple domains. 
From the 19 available datasets, we exclude those used for fine-tuning the tested retrievers (e.g., MSMARCO, Natural Questions, FEVER, HotpotQA, CQADupStack), as well as those that are no longer accessible (e.g., TREC-News, Robust04, Signal-1M, BioASQ), following \citet{khramtsova2023selecting}. This leaves us with 10 datasets for evaluation. Each dataset consists of a document corpus and query-document pairs. In our experiment, we define the corpus $\mathcal{C}$ as the set of documents with at least one annotated relevant query, ensuring a quantitative evaluation.

\textbf{Baselines.}\label{sec:exp-prediction-baelines}
We compare our method with three baselines: (i) \textbf{Layerwise} \citep{izacardunsupervised}: unsupervised textual OOD detection via layerwise anomaly scores (e.g., negative cosine similarity), (ii) \textbf{IPQ} \citep{chen2023continual}: incremental production quantization with clustering, and (iii) \textbf{GenQuery} \citep{khramtsova2024leveraging}: zero-shot ranking using pseudo-questions generated by large language models. Implementation details for each baseline are provided within each respective experiment.

\textbf{Hyperparameters.} 
To calculate the gradient norm for each document $d$, we set the dropout rate to 0.02 and the number of positives ($p$) to 8. We use four negative samples ($n$) to reduce the computational cost.
For OOD detection, we use the average gradient norm of 3,000 in-domain Natural Questions (NQ) documents \citep{kwiatkowski2019natural} as the reference threshold, since all test retrievers are already trained on NQ. Documents with gradient norms exceeding this average are classified as OOD. We set the OOD corpus prediction threshold ($\gamma$) to 0.5. We determine this value empirically based on preliminary experiments. We conduct an extensive ablation study as described in Appendix~\ref{app:ablation}, exhibiting the robustness of our approach.

\begin{table*}[t!]
\begin{center}
\renewcommand{\arraystretch}{0.98}
\resizebox{\linewidth}{!}{
    \begin{tabular}{p{1.5cm} >{\centering\arraybackslash}p{1.7cm} >{\centering\arraybackslash}p{1.9cm} >{\centering\arraybackslash}p{1.7cm} >{\centering\arraybackslash}p{1.7cm} >{\centering\arraybackslash}p{1.7cm} >{\centering\arraybackslash}p{1.7cm} >{\centering\arraybackslash}p{1.7cm} >{\centering\arraybackslash}p{1.7cm} >{\centering\arraybackslash}p{1.7cm} >{\centering\arraybackslash}p{1.7cm}} 
    \toprule
    \textbf{Method} & ArguAna & C-FEVER & DBPedia & FiQA & NFCorpus & Quora & Scidocs & SciFact & COVID & Touch\'e \\ \midrule
    Layerwise & \textbf{99.36} & 79.14 & 85.45 & 77.25 & 28.88 & \textbf{99.97} & 59.06 & 98.39 & 18.53 & 97.16 \\
    IPQ & 99.00 & 79.14 & 85.45 & 77.25 & 28.88 & \textbf{99.97} & 59.06 & 98.39 & 18.53 & 97.16 \\
    GenQuery & 99.00 & 79.14 & 85.45 & 77.25 & 28.88 & \textbf{99.97} & 59.06 & 98.15 & 18.53 & 97.16 \\
    Ours & \textbf{99.36} & \textbf{82.55} & \textbf{89.73} & \textbf{78.51} & \textbf{35.73} & \textbf{99.97} & \textbf{69.63} & \textbf{99.73} & \textbf{21.52} & \textbf{98.40} \\ \midrule
    \textcolor{mygray}{Oracle} & \textcolor{mygray}{99.43} & \textcolor{mygray}{82.92} & \textcolor{mygray}{89.73} & \textcolor{mygray}{83.25} & \textcolor{mygray}{35.73} & \textcolor{mygray}{99.97} & \textcolor{mygray}{69.98} & \textcolor{mygray}{99.73} & \textcolor{mygray}{21.52} & \textcolor{mygray}{99.23} \\
    \bottomrule
\end{tabular}
}
\caption{Results of zero-shot retriever selection in terms of Recall@100 scores of the retriever selected by each OOD method. The \textit{oracle} is the upper bound, indicating the performance of the actual best retriever per dataset. 
}
\vspace{-0.5cm}
\label{tab:exp-selection}
\end{center}
\end{table*}

\subsection{Detection of OOD Documents}\label{sec:exp-ood-doc}

This task aims to detect OOD documents from a new document corpus $\mathcal{C}$. 
Our method selects OOD documents where GradNormIR exceeds the threshold as described in Section~\ref{sec:exp-setup}. For other baselines, we rank the documents in descending order by their OOD scores as described below, and then select the same number of top-ranked documents as GradNormIR for fairness. Finally, we compare these detected OOD documents using their retrieval rate using query-document pairs in the dataset.

\textbf{Evaluation Metric.} To evaluate the OOD document detection, we use the document retrieval rate (DRR). As described in Section~\ref{sec:ood-definition}, the effectiveness of an approach can be measured by how poorly detected OOD documents are retrieved by relevant queries. 
For each dataset, we organize annotations as $\{d_i, Q_{d_i}\}_{i=1}^N$, where $Q_{d_i}$ represents the set of relevant queries for each document $d_i$. DRR is then calculated as
\begin{equation}\label{eq:acc}
\text{DRR} = \frac{\sum_{d_i \in \mathcal{C}}\sum_{q_{d_i} \in Q_{d_i}} \mathds{1} \{d_i \in D^+(q_{d_i})\}}{\sum_{d_i \in \mathcal{C}} |Q_{d_i}|}, 
\end{equation}
where $\mathds{1}$ is an indicator function that returns 1 if $d_i$ appears in the top-$k$ retrieval results $D^+(q_{d_i})$ (with $k=100$), and 0 otherwise. 
Lower DRR values indicate that the identified OOD documents are indeed retrieved less frequently, validating effective OOD detection.

\textbf{Implementations of Baselines.} For each baseline, we first compute the OOD score of each document $d$ as follows: (i) \textbf{Layerwise}: we compute the negative cosine similarity between latent vectors of $d$ and in-domain documents across all layers and aggregate them to produce a final OOD score for $d$. (ii) \textbf{IPQ} creates quantization codebooks from $\mathcal{C}$ to get centroids. We quantize all representations to generate centroids and use the average Euclidean distance between the quantized representation and the centroids as the OOD score of $d$. (iii) \textbf{GenQuery}: we generate a pseudo-question $\hat{q}$ for $d$ using Llama3.1-8B. We then use the rank of $d$ in the retrieval results of $\hat{q}$ as the OOD score. 

\subsubsection{Results}
Table~\ref{tab:exp-prediction} presents the results of OOD document detection across 10 datasets of the BEIR benchmark. Our GradNormIR consistently outperforms the baselines, achieving the lowest average DRR on all tested retrievers. 
Notably, GradNormIR substantially reduces DRR on DBPedia-Entity and Scidocs, achieving reductions of 28.18 and 15.36, respectively, for BGE.

In the baselines, the detected OOD documents often show unexpectedly higher retrieval rates than the average DRR of all documents, indicating wrong detection. For instance, GenQuery in DBPedia-Entity shows significant increases across all retrievers, although it achieves the best performance on Quora for Contriever, E5, and GTE. 
Also, in Climate-FEVER, GenQuery increases for BGE, E5, and GTE. This may be because these documents are also out-of-domain to the LLM. 
Typically, IPQ and Layerwise baselines show the lowest DRRs in some cases, but their performance fluctuates up and down, indicating low robustness.

Overall, GradNormIR consistently shows lower document retrieval rates for detected OOD documents, demonstrating that it accurately identifies OOD documents across datasets. We further evaluate the OOD documents ratio, \(r(\mathcal{C})\) in Section~\ref{eq:ood-corpus} in the following experiments.

\subsection{Best Retriever Selection}

This task predicts the most suitable dense retriever from a set of retrievers given a corpus $\mathcal{C}$, i.e., it selects the retriever with the highest generalizability for the given $\mathcal{C}$ using the OOD detection method. This task shows that our approach is helpful for selecting not only when the retrievers are updated, but also which one is the best in the stream of corpora. 

\textbf{Setup.} We select one of four retrievers (as described in Section~\ref{sec:exp-prediction-retrievers}), choosing the one that has the lowest OOD document ratio, \(r(\mathcal{C})\). Specifically, given a test dataset including $\mathcal{C}$ and query-document pairs, we calculate $r(\mathcal{C})$ for each retriever. Next, we select the retriever with the lowest $r(\mathcal{C})$. Then, we evaluate the selected retriever on the query-document pairs. 
For each dataset, we report the Recall@100 performance of the retriever selected by each baseline. 

\textbf{Implementations of Baselines.} To calculate  $r(\mathcal{C})$, we first compute the OOD score of the in-domain NQ documents for each baseline in the same way as in Section~\ref{sec:exp-ood-doc}. Then, we calculate the ratio of documents with an OOD score greater than the median. In this way, we can compute $r(\mathcal{C})$ of each baseline. 

\subsubsection{Results}
Table~\ref{tab:exp-selection} presents the Recall@100 performance of the selected retriever by each baseline.  The \textit{oracle} row shows the performance of the actual optimal retriever on each dataset. 
The retriever chosen by GradNormIR consistently achieves the highest performance across datasets. While GradNormIR does not always select the top-performing retriever (e.g., BGE for ArguAna, GTE for FiQA), it consistently identifies the second-best retriever, indicating robust generalizability estimation. These results show that GradNormIR is highly effective in selecting the most appropriate retriever based solely on the given document corpus, even before any queries are introduced.

\subsection{Continual Updates}

The goal of this task is to update the retriever only when an OOD corpus is given, balancing performance stability and computational cost in evolving corpora. 

\textbf{Setup.} 
We simulate the sequential streaming of a corpus using datasets of the BEIR coming in alphabetical order. This experimental setup mirrors approaches in continual learning research, such as the \citep{gelightweight}, which sequentially combines several heterogeneous image classification datasets to evaluate models' abilities to adapt to diverse and evolving tasks. Specifically, in session $S_1$, the Arguana corpus is given, in session $S_2$, the Climate-FEVER corpus is given, and so on. 
We continually update Contriever using RecAdam optimizer \citep{chen-etal-2020-recall}, widely employed to mitigate the language model's catastrophic forgetting. In session $S_t$, we update the current retriever with a given corpus. We then build a retrieval index using corpus from $S_1$ to $S_t$. Finally, we evaluate the retriever with queries from $S_1$ to $S_t$. For training details, please refer to Appendix~\ref{app:experimental_details}.

\textbf{Implementations of Baselines.} We test three types of baselines: (i) Zero-shot: the retriever remains fixed with no further updates. (ii) Selective: the retriever is updated only when a newly given corpus is determined as an OOD corpus. (iii) Na\"ive: the retriever is updated whenever a new corpus is given, common in continual learning. For selective retraining, in each session $S_t$, we decide whether to update the retriever and use the most recently updated retriever to build an index using corpus from $S_1$ to $S_t$. We evaluate different update strategies from the four baselines. In GradNormIR, we update the retriever when a corpus is OOD, in total $N$ times ($N=6$). For the other retraining methods, the retriever undergoes the same $N$ updates with the corpora of the highest OOD ratios for fairness.

\textbf{Metrics.} We compute the average Recall@100 in each session \( S_t \), computed as the mean Recall@100 of the datasets from \( S_1 \) to \( S_t \). We report the relative performance with respect to an upper bound per dataset, since each dataset has different levels of difficulty. The upper bound of each dataset is the Recall@100 value of the retriever fine-tuned only with the dataset. 

\begin{figure}[t]
    \includegraphics[width=0.49\textwidth]
    {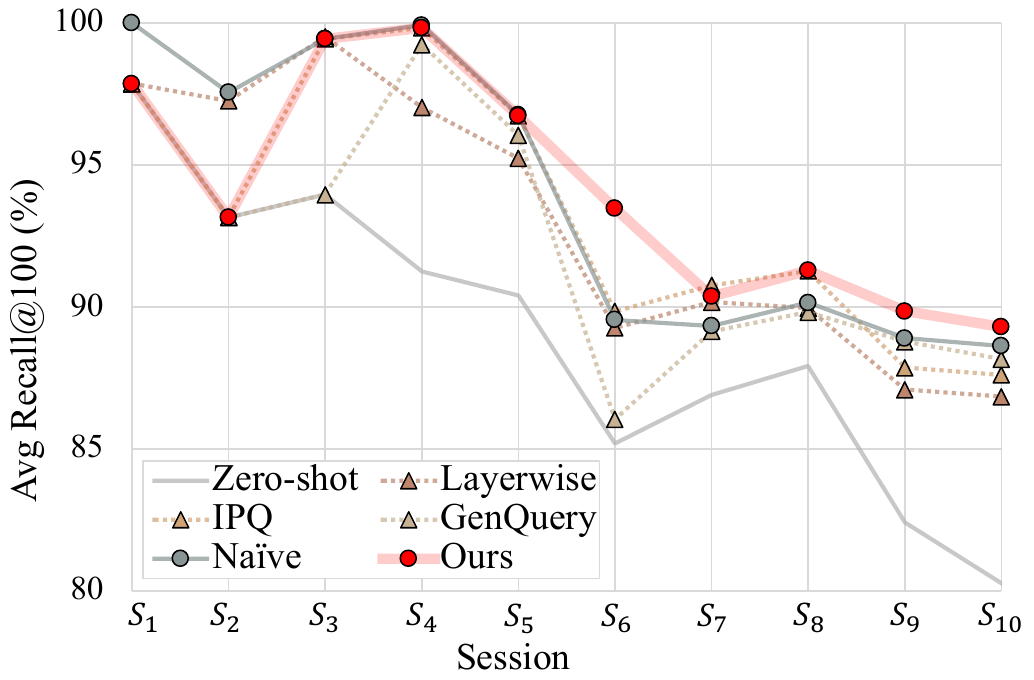}
    \caption{
    Average of Recall@100 across $S_1$ to $S_t$ with respect to the upper bound for each dataset using a single-trained retriever. Although the trend decreases due to the expanding document corpus over sessions, performance remains robust with continual updates.
    }
    \label{fig:continual}
    \vspace{-0.5cm}
\end{figure}

\subsubsection{Results}

Figure~\ref{fig:continual} illustrates the retrieval results of different continual update strategies over the sessions. Overall, performance degrades as the sessions progress. This occurs because as the number of documents increases, the corpus expands, making it more difficult to retrieve the correct documents. Thus, the performance of the Zero-shot baseline quite drops to around 80. However, with continual updates, the other baselines maintain stable performance around 90, preventing catastrophic forgetting.

Initially, GradNormIR exhibits lower performance in $S_1$ and $S_2$, since the retriever is not updated.
Nonetheless, it does not show significant degradation afterward, maintaining the retriever's accuracy by retraining in later sessions. Starting from session $S_6$, GradNormIR achieves the highest average performance among all baselines. Notably, in $S_6$, GradNormIR outperforms even the Na\"ive baseline, which retrains the retriever in every session. This indicates that unnecessary continual retraining can negatively impact retrieval performance, highlighting the importance of selective retraining. The performance gap persists until the final session, demonstrating the efficiency and effectiveness of GradNormIR's selective retraining.

Conversely, all other selective baselines exhibit lower performance than Na\"ive baseline. For instance, Layerwise displays robust performance in the earlier sessions, but it shows persistent performance degradation in later sessions since it is not trained on the OOD corpus in $S_4$ and $S_6$. This suggests that selective retraining only with the OOD corpus can ensure the maintenance of retriever performance in evolving corpora.

\begin{table}[t!]
\begin{center}
\renewcommand{\arraystretch}{1}
\resizebox{\linewidth}{!}{
    \begin{tabular}{>{\centering\arraybackslash}p{1.6cm} >{\centering\arraybackslash}p{1.8cm} >{\centering\arraybackslash}p{1.5cm} >{\centering\arraybackslash}p{1.5cm} >{\centering\arraybackslash}p{1.5cm} >{\centering\arraybackslash}p{1.5cm}} 
    \toprule
    \multirow{2}{*}{\textbf{Hard Neg}} & \multirow{2}{*}{\textbf{Dropout}} & \multicolumn{4}{c}{DRR ($\downarrow$)} \\ \cmidrule(lr){3-6}
    & & BGE & Cont & E5 & GTE \\ \midrule
    & \cmark & 67.68 & 64.19 & 67.45 & \textbf{68.07} \\ 
    \cmark & & 65.79 & 62.41 & \textbf{65.58} & 70.75 \\ 
    \cmark & \cmark & \textbf{65.03} & \textbf{61.46} & 65.59 & 68.62 \\ 
    \bottomrule
\end{tabular}
}
\caption{Ablation study on the impact of dropout for document queries and the use of hard negatives.
}
\vspace{-0.7cm}
\label{tab:exp-ablation}
\end{center}
\end{table}

\subsection{Ablation Study}

We evaluate the impact of the dropout and the use of hard negatives. Table~\ref{tab:exp-ablation} displays the results of the average DRR in OOD document detection. When both dropout and hard negatives are applied, the model achieves the best performance, particularly for the BGE and Contriever. For E5, hard negatives contribute to an increase in DRR, while dropout also proves effective. Conversely, for GTE, hard negatives enhance performance, whereas dropout leads to performance degradation. This suggests that the optimal setting may vary depending on the chosen retriever. Nonetheless, even in these two cases, both hard negatives and dropout yield reasonable performance, indicating the robustness of GradNormIR to hyperparameter choices. Additional ablation experiments are provided in Appendix~\ref{app:ablation}.

\section{Conclusion}

We introduced the novel task of predicting an OOD corpus for dense retrievers before indexing, addressing a critical challenge for maintaining robust retrieval performance in dynamically evolving corpora. To tackle this problem, we proposed \textbf{GradNormIR}, an unsupervised method leveraging gradient norms of the contrastive loss to proactively detect OOD documents. By employing novel sampling strategies, including document-to-document retrieval with carefully selected positive and hard negative samples, GradNormIR can effectively predict whether a corpus might pose retrieval challenges even before query collection. Our method enables timely selection of the most suitable retriever or updates to an existing retriever, thereby ensuring robust performance in evolving document environments.

An interesting direction for future research would involve online OOD document detection, where documents arrive continuously rather than as a complete corpus. Such an extension could enhance the practical applicability of retrievers in real-time retrieval scenarios.

\section*{Limitations}

While our work effectively demonstrates the importance and efficacy of proactively detecting OOD documents and corpora, there are several limitations that warrant consideration. First, GradNormIR operates solely at the document level without considering potential future queries explicitly. Consequently, retrieval performance may still degrade if unexpected queries significantly differ from the assumed document representations. Incorporating query-aware mechanisms could further enhance robustness against such scenarios. Also, our approach relies on document-to-document similarity computations, which may pose scalability challenges when applied to extremely large corpora. In practical large-scale deployments, it might be necessary to implement efficient indexing strategies or divide corpora into smaller subsets for manageable processing. We validate these strategies in Appendix, so please refer to it.

\section*{Ethics Statement}

This research does not raise ethical concerns, as it primarily addresses technical advancements in information retrieval models and their evaluation methodologies. Our methods focus exclusively on improving retrieval performance using non-sensitive, general-purpose datasets and do not involve handling any personal, confidential, or ethically sensitive information.

\section*{Acknowledgments}
We sincerely thank Heeseung Yun, Sangwoo Moon, and other anonymous reviewers for their valuable comments. 
This work was supported by 
supported by Samsung SDS, 
Korea Radio Promotion Association (Development of Intelligent Docent Service for Information-Disadvantaged Groups), 
Institute of Information \& Communications Technology Planning \& Evaluation (IITP) grant funded by the Korea government (MSIT) (No.~RS-2019-II191082, SW StarLab), 
Institute of Information \& communications Technology Planning \& Evaluation (IITP) grant funded by the Korea government (MSIT) (No.~RS-2022-II220156, Fundamental research on continual meta-learning for quality enhancement of casual videos and their 3D metaverse transformation), 
and the National Research Foundation of Korea (NRF) grant funded by the Korea government (MSIT) (No.~2023R1A2C2005573). 
Gunhee Kim is the corresponding author. 



\begin{thebibliography}{39}
\providecommand{\natexlab}[1]{#1}

\bibitem[{Bajaj et~al.(2016)Bajaj, Campos, Craswell, Deng, Gao, Liu, Majumder, McNamara, Mitra, Nguyen et~al.}]{bajaj2016ms}
Payal Bajaj, Daniel Campos, Nick Craswell, Li~Deng, Jianfeng Gao, Xiaodong Liu, Rangan Majumder, Andrew McNamara, Bhaskar Mitra, Tri Nguyen, et~al. 2016.
\newblock Ms marco: A human generated machine reading comprehension dataset.
\newblock \emph{arXiv preprint arXiv:1611.09268}.

\bibitem[{Besta et~al.(2024)Besta, Kubicek, Niggli, Gerstenberger, Weitzendorf, Chi, Iff, Gajda, Nyczyk, M{\"u}ller et~al.}]{besta2024multi}
Maciej Besta, Ales Kubicek, Roman Niggli, Robert Gerstenberger, Lucas Weitzendorf, Mingyuan Chi, Patrick Iff, Joanna Gajda, Piotr Nyczyk, J{\"u}rgen M{\"u}ller, et~al. 2024.
\newblock Multi-head rag: Solving multi-aspect problems with llms.
\newblock \emph{arXiv preprint arXiv:2406.05085}.

\bibitem[{Cai et~al.(2023)Cai, Bi, Fan, Guo, Chen, and Cheng}]{cai2023l2r}
Yinqiong Cai, Keping Bi, Yixing Fan, Jiafeng Guo, Wei Chen, and Xueqi Cheng. 2023.
\newblock L2r: Lifelong learning for first-stage retrieval with backward-compatible representations.
\newblock In \emph{Proceedings of the 32nd ACM International Conference on Information and Knowledge Management}, pages 183--192.

\bibitem[{Chen et~al.(2023)Chen, Zhang, Guo, de~Rijke, Chen, Fan, and Cheng}]{chen2023continual}
Jiangui Chen, Ruqing Zhang, Jiafeng Guo, Maarten de~Rijke, Wei Chen, Yixing Fan, and Xueqi Cheng. 2023.
\newblock Continual learning for generative retrieval over dynamic corpora.
\newblock In \emph{Proceedings of the 32nd ACM International Conference on Information and Knowledge Management}, pages 306--315.

\bibitem[{Chen et~al.(2024)Chen, Xiao, Zhang, Luo, Lian, and Liu}]{chen2024bge}
Jianlv Chen, Shitao Xiao, Peitian Zhang, Kun Luo, Defu Lian, and Zheng Liu. 2024.
\newblock Bge m3-embedding: Multi-lingual, multi-functionality, multi-granularity text embeddings through self-knowledge distillation.
\newblock \emph{arXiv preprint arXiv:2402.03216}.

\bibitem[{Chen et~al.(2020)Chen, Hou, Cui, Che, Liu, and Yu}]{chen-etal-2020-recall}
Sanyuan Chen, Yutai Hou, Yiming Cui, Wanxiang Che, Ting Liu, and Xiangzhan Yu. 2020.
\newblock \href {https://doi.org/10.18653/v1/2020.emnlp-main.634} {Recall and learn: Fine-tuning deep pretrained language models with less forgetting}.
\newblock In \emph{Proceedings of the 2020 Conference on Empirical Methods in Natural Language Processing (EMNLP)}, pages 7870--7881, Online. Association for Computational Linguistics.

\bibitem[{Chen et~al.(2022)Chen, Zhang, Lu, Bendersky, and Najork}]{chen2022out}
Tao Chen, Mingyang Zhang, Jing Lu, Michael Bendersky, and Marc Najork. 2022.
\newblock Out-of-domain semantics to the rescue! zero-shot hybrid retrieval models.
\newblock In \emph{European Conference on Information Retrieval}, pages 95--110.

\bibitem[{Darrin et~al.(2024)Darrin, Staerman, Gomes, Cheung, Piantanida, and Colombo}]{darrin2024unsupervised}
Maxime Darrin, Guillaume Staerman, Eduardo Dadalto~C{\^a}mara Gomes, Jackie~CK Cheung, Pablo Piantanida, and Pierre Colombo. 2024.
\newblock Unsupervised layer-wise score aggregation for textual ood detection.
\newblock In \emph{Proceedings of the AAAI Conference on Artificial Intelligence}, volume~38, pages 17880--17888.

\bibitem[{Gao et~al.(2021)Gao, Yao, and Chen}]{gao2021simcse}
Tianyu Gao, Xingcheng Yao, and Danqi Chen. 2021.
\newblock Simcse: Simple contrastive learning of sentence embeddings.
\newblock \emph{arXiv preprint arXiv:2104.08821}.

\bibitem[{Ge et~al.(2023)Ge, Li, Wu, Xu, Jones, Rios, Fostiropoulos, Huang, Murdock, Sahin et~al.}]{gelightweight}
Yunhao Ge, Yuecheng Li, Di~Wu, Ao~Xu, Adam~M Jones, Amanda~Sofie Rios, Iordanis Fostiropoulos, Po-Hsuan Huang, Zachary~William Murdock, Gozde Sahin, et~al. 2023.
\newblock Lightweight learner for shared knowledge lifelong learning.
\newblock In \emph{ICML Workshop on Localized Learning (LLW)}.

\bibitem[{Huang et~al.(2021)Huang, Geng, and Li}]{huang2021importance}
Rui Huang, Andrew Geng, and Yixuan Li. 2021.
\newblock On the importance of gradients for detecting distributional shifts in the wild.
\newblock \emph{Advances in Neural Information Processing Systems}, 34:677--689.

\bibitem[{Izacard et~al.(2022)Izacard, Caron, Hosseini, Riedel, Bojanowski, Joulin, and Grave}]{izacardunsupervised}
Gautier Izacard, Mathilde Caron, Lucas Hosseini, Sebastian Riedel, Piotr Bojanowski, Armand Joulin, and Edouard Grave. 2022.
\newblock Unsupervised dense information retrieval with contrastive learning.
\newblock \emph{Transactions on Machine Learning Research}.

\bibitem[{Jeong et~al.(2022)Jeong, Baek, Cho, Hwang, and Park}]{jeong2022augmenting}
Soyeong Jeong, Jinheon Baek, Sukmin Cho, Sung~Ju Hwang, and Jong~C Park. 2022.
\newblock Augmenting document representations for dense retrieval with interpolation and perturbation.
\newblock In \emph{Proceedings of the 60th Annual Meeting of the Association for Computational Linguistics (Volume 2: Short Papers)}, pages 442--452.

\bibitem[{Karpukhin et~al.(2020)Karpukhin, Oguz, Min, Lewis, Wu, Edunov, Chen, and Yih}]{karpukhin2020dense}
Vladimir Karpukhin, Barlas Oguz, Sewon Min, Patrick Lewis, Ledell Wu, Sergey Edunov, Danqi Chen, and Wen-tau Yih. 2020.
\newblock Dense passage retrieval for open-domain question answering.
\newblock In \emph{Proceedings of the 2020 Conference on Empirical Methods in Natural Language Processing (EMNLP)}, pages 6769--6781.

\bibitem[{Kasela et~al.(2024)Kasela, Pasi, Perego, and Tonellotto}]{kasela2024desire}
Pranav Kasela, Gabriella Pasi, Raffaele Perego, and Nicola Tonellotto. 2024.
\newblock Desire-me: Domain-enhanced supervised information retrieval using mixture-of-experts.
\newblock In \emph{European Conference on Information Retrieval}, pages 111--125.

\bibitem[{Khramtsova et~al.(2023)Khramtsova, Zhuang, Baktashmotlagh, Wang, and Zuccon}]{khramtsova2023selecting}
Ekaterina Khramtsova, Shengyao Zhuang, Mahsa Baktashmotlagh, Xi~Wang, and Guido Zuccon. 2023.
\newblock Selecting which dense retriever to use for zero-shot search.
\newblock In \emph{Proceedings of the Annual International ACM SIGIR Conference on Research and Development in Information Retrieval in the Asia Pacific Region}, pages 223--233.

\bibitem[{Khramtsova et~al.(2024)Khramtsova, Zhuang, Baktashmotlagh, and Zuccon}]{khramtsova2024leveraging}
Ekaterina Khramtsova, Shengyao Zhuang, Mahsa Baktashmotlagh, and Guido Zuccon. 2024.
\newblock Leveraging llms for unsupervised dense retriever ranking.
\newblock In \emph{Proceedings of the 47th International ACM SIGIR Conference on Research and Development in Information Retrieval}, pages 1307--1317.

\bibitem[{Kim et~al.(2024)Kim, Ko, and Kim}]{kim-etal-2024-dynamicer}
Jinyoung Kim, Dayoon Ko, and Gunhee Kim. 2024.
\newblock \href {https://doi.org/10.18653/v1/2024.emnlp-main.762} {{D}ynamic{ER}: Resolving emerging mentions to dynamic entities for {RAG}}.
\newblock In \emph{Proceedings of the 2024 Conference on Empirical Methods in Natural Language Processing}, pages 13752--13770, Miami, Florida, USA. Association for Computational Linguistics.

\bibitem[{Ko et~al.(2024)Ko, Kim, Choi, and Kim}]{ko-etal-2024-growover}
Dayoon Ko, Jinyoung Kim, Hahyeon Choi, and Gunhee Kim. 2024.
\newblock \href {https://doi.org/10.18653/v1/2024.acl-long.181} {{G}row{OVER}: How can {LLM}s adapt to growing real-world knowledge?}
\newblock In \emph{Proceedings of the 62nd Annual Meeting of the Association for Computational Linguistics (Volume 1: Long Papers)}, pages 3282--3308, Bangkok, Thailand. Association for Computational Linguistics.

\bibitem[{Kwiatkowski et~al.(2019)Kwiatkowski, Palomaki, Redfield, Collins, Parikh, Alberti, Epstein, Polosukhin, Devlin, Lee et~al.}]{kwiatkowski2019natural}
Tom Kwiatkowski, Jennimaria Palomaki, Olivia Redfield, Michael Collins, Ankur Parikh, Chris Alberti, Danielle Epstein, Illia Polosukhin, Jacob Devlin, Kenton Lee, et~al. 2019.
\newblock Natural questions: A benchmark for question answering research.
\newblock \emph{Transactions of the Association for Computational Linguistics}, 7:452--466.

\bibitem[{Lee et~al.(2024)Lee, Soldaini, Cohan, Seo, and Lo}]{lee2024routerretriever}
Hyunji Lee, Luca Soldaini, Arman Cohan, Minjoon Seo, and Kyle Lo. 2024.
\newblock Routerretriever: Exploring the benefits of routing over multiple expert embedding models.
\newblock \emph{arXiv preprint arXiv:2409.02685}.

\bibitem[{Lewis et~al.(2020)Lewis, Perez, Piktus, Petroni, Karpukhin, Goyal, K{\"u}ttler, Lewis, Yih, Rockt{\"a}schel et~al.}]{lewis2020retrieval}
Patrick Lewis, Ethan Perez, Aleksandra Piktus, Fabio Petroni, Vladimir Karpukhin, Naman Goyal, Heinrich K{\"u}ttler, Mike Lewis, Wen-tau Yih, Tim Rockt{\"a}schel, et~al. 2020.
\newblock Retrieval-augmented generation for knowledge-intensive nlp tasks.
\newblock \emph{Advances in Neural Information Processing Systems}, 33:9459--9474.

\bibitem[{Leys et~al.(2013)Leys, Ley, Klein, Bernard, and Licata}]{leys2013detecting}
Christophe Leys, Christophe Ley, Olivier Klein, Philippe Bernard, and Laurent Licata. 2013.
\newblock Detecting outliers: Do not use standard deviation around the mean, use absolute deviation around the median.
\newblock \emph{Journal of experimental social psychology}, 49(4):764--766.

\bibitem[{Li et~al.(2023{\natexlab{a}})Li, Rawat, Zaheer, Wang, Lukasik, Veit, Yu, and Kumar}]{li2023large}
Daliang Li, Ankit~Singh Rawat, Manzil Zaheer, Xin Wang, Michal Lukasik, Andreas Veit, Felix Yu, and Sanjiv Kumar. 2023{\natexlab{a}}.
\newblock Large language models with controllable working memory.
\newblock In \emph{Findings of the Association for Computational Linguistics: ACL 2023}, pages 1774--1793.

\bibitem[{Li et~al.(2023{\natexlab{b}})Li, Zhang, Zhang, Long, Xie, and Zhang}]{li2023towards}
Zehan Li, Xin Zhang, Yanzhao Zhang, Dingkun Long, Pengjun Xie, and Meishan Zhang. 2023{\natexlab{b}}.
\newblock Towards general text embeddings with multi-stage contrastive learning.
\newblock \emph{arXiv preprint arXiv:2308.03281}.

\bibitem[{Liu et~al.(2024)Liu, Zhang, Guo, de~Rijke, Fan, and Cheng}]{liu2024robust}
Yu-An Liu, Ruqing Zhang, Jiafeng Guo, Maarten de~Rijke, Yixing Fan, and Xueqi Cheng. 2024.
\newblock Robust neural information retrieval: An adversarial and out-of-distribution perspective.
\newblock \emph{arXiv preprint arXiv:2407.06992}.

\bibitem[{Ni et~al.(2021)Ni, Qu, Lu, Dai, Ábrego, Ma, Zhao, Luan, Hall, Chang, and Yang}]{ni2021largedualencodersgeneralizable}
Jianmo Ni, Chen Qu, Jing Lu, Zhuyun Dai, Gustavo~Hernández Ábrego, Ji~Ma, Vincent~Y. Zhao, Yi~Luan, Keith~B. Hall, Ming-Wei Chang, and Yinfei Yang. 2021.
\newblock \href {https://arxiv.org/abs/2112.07899} {Large dual encoders are generalizable retrievers}.
\newblock \emph{Preprint}, arXiv:2112.07899.

\bibitem[{Petroni et~al.()Petroni, Lewis, Piktus, Rockt{\"a}schel, Wu, Miller, and Riedel}]{petronicontext}
Fabio Petroni, Patrick Lewis, Aleksandra Piktus, Tim Rockt{\"a}schel, Yuxiang Wu, Alexander~H Miller, and Sebastian Riedel.
\newblock How context affects language models' factual predictions.
\newblock In \emph{Automated Knowledge Base Construction}.

\bibitem[{Ramos et~al.(2003)}]{ramos2003using}
Juan Ramos et~al. 2003.
\newblock Using tf-idf to determine word relevance in document queries.
\newblock In \emph{Proceedings of the first instructional conference on machine learning}, volume 242, pages 29--48. Citeseer.

\bibitem[{Robertson et~al.(2009)Robertson, Zaragoza et~al.}]{robertson2009probabilistic}
Stephen Robertson, Hugo Zaragoza, et~al. 2009.
\newblock The probabilistic relevance framework: Bm25 and beyond.
\newblock \emph{Foundations and Trends{\textregistered} in Information Retrieval}, 3(4):333--389.

\bibitem[{Thakur et~al.(2021)Thakur, Reimers, R{\"u}ckl{\'e}, Srivastava, and Gurevych}]{thakur2beir}
Nandan Thakur, Nils Reimers, Andreas R{\"u}ckl{\'e}, Abhishek Srivastava, and Iryna Gurevych. 2021.
\newblock Beir: A heterogeneous benchmark for zero-shot evaluation of information retrieval models.
\newblock In \emph{Thirty-fifth Conference on Neural Information Processing Systems Datasets and Benchmarks Track}.

\bibitem[{Wang et~al.(2021)Wang, Thakur, Reimers, and Gurevych}]{wang2021gpl}
Kexin Wang, Nandan Thakur, Nils Reimers, and Iryna Gurevych. 2021.
\newblock Gpl: Generative pseudo labeling for unsupervised domain adaptation of dense retrieval.
\newblock \emph{arXiv preprint arXiv:2112.07577}.

\bibitem[{Wang et~al.(2022)Wang, Yang, Huang, Jiao, Yang, Jiang, Majumder, and Wei}]{wang2022text}
Liang Wang, Nan Yang, Xiaolong Huang, Binxing Jiao, Linjun Yang, Daxin Jiang, Rangan Majumder, and Furu Wei. 2022.
\newblock Text embeddings by weakly-supervised contrastive pre-training.
\newblock \emph{arXiv preprint arXiv:2212.03533}.

\bibitem[{Wang et~al.(2024)Wang, Yang, Huang, Yang, Majumder, and Wei}]{wang2024multilingual}
Liang Wang, Nan Yang, Xiaolong Huang, Linjun Yang, Rangan Majumder, and Furu Wei. 2024.
\newblock Multilingual e5 text embeddings: A technical report.
\newblock \emph{arXiv preprint arXiv:2402.05672}.

\bibitem[{Xiao et~al.(2023)Xiao, Liu, Zhang, and Muennighoff}]{bge_embedding}
Shitao Xiao, Zheng Liu, Peitian Zhang, and Niklas Muennighoff. 2023.
\newblock \href {https://arxiv.org/abs/2309.07597} {C-pack: Packaged resources to advance general chinese embedding}.
\newblock \emph{Preprint}, arXiv:2309.07597.

\bibitem[{Xie et~al.(2024)Xie, Odonnat, Feofanov, Redko, Zhang, and An}]{xie2024characterising}
Renchunzi Xie, Ambroise Odonnat, Vasilii Feofanov, Ievgen Redko, Jianfeng Zhang, and Bo~An. 2024.
\newblock Characterising gradients for unsupervised accuracy estimation under distribution shift.
\newblock \emph{arXiv preprint arXiv:2401.08909}.

\bibitem[{Yang et~al.(2018)Yang, Qi, Zhang, Bengio, Cohen, Salakhutdinov, and Manning}]{yang2018hotpotqa}
Zhilin Yang, Peng Qi, Saizheng Zhang, Yoshua Bengio, William Cohen, Ruslan Salakhutdinov, and Christopher~D Manning. 2018.
\newblock Hotpotqa: A dataset for diverse, explainable multi-hop question answering.
\newblock In \emph{Proceedings of the 2018 Conference on Empirical Methods in Natural Language Processing}, pages 2369--2380.

\bibitem[{Yu et~al.(2022)Yu, Xiong, Sun, Zhang, and Overwijk}]{yu2022coco}
Yue Yu, Chenyan Xiong, Si~Sun, Chao Zhang, and Arnold Overwijk. 2022.
\newblock Coco-dr: Combating distribution shift in zero-shot dense retrieval with contrastive and distributionally robust learning.
\newblock In \emph{Proceedings of the 2022 Conference on Empirical Methods in Natural Language Processing}, pages 1462--1479.

\bibitem[{Zhan et~al.(2021)Zhan, Mao, Liu, Guo, Zhang, and Ma}]{zhan2021optimizing}
Jingtao Zhan, Jiaxin Mao, Yiqun Liu, Jiafeng Guo, Min Zhang, and Shaoping Ma. 2021.
\newblock Optimizing dense retrieval model training with hard negatives.
\newblock In \emph{Proceedings of the 44th International ACM SIGIR Conference on Research and Development in Information Retrieval}, pages 1503--1512.

\end{thebibliography}

\newpage

\appendix
\newpage
\section{Experiment Details}\label{app:experimental_details}

\textbf{Models.} 
In experiments, we use four dense retrievers: BGE-large-en-v1.5, unsupervised Contriever, multilingual E5 large, and GTE-base. In hugging face, the model names are \textit{BAAI/bge-large-en-v1.5}, \textit{facebook/contriever}, \textit{intfloat/multilingual-e5-large}, and \textit{thenlper/gte-base}, respectively. 

\textbf{GradNormIR.} For contrastive loss temperature $\tau$, we use 0.05 for all baselines but 0.01 for e5. The probability distribution is skewed for E5, as noted in hugging face; setting the temperature to 0.05 does not make the model compute contrastive loss effectively. 

For positive and negative sampling, we sample 8 positive samples ($p$) and four negatives ($n$) for each positive. As sampling 4 negatives per positive is traditional, we follow previous work.

For the dropout, we use 0.02, which means 2 percent of tokens are masked to zero. For the other experimental setups, we follow the default values of BAAI Flagembeddings\footnote{https://github.com/FlagOpen/FlagEmbedding}. 

We conduct ablation studies on the impact of the number of positive samples and dropout rate in Section~\ref{sec:ablation-dropout}.

\begin{figure*}[b!]
\centering
\includegraphics[width=0.8\textwidth]{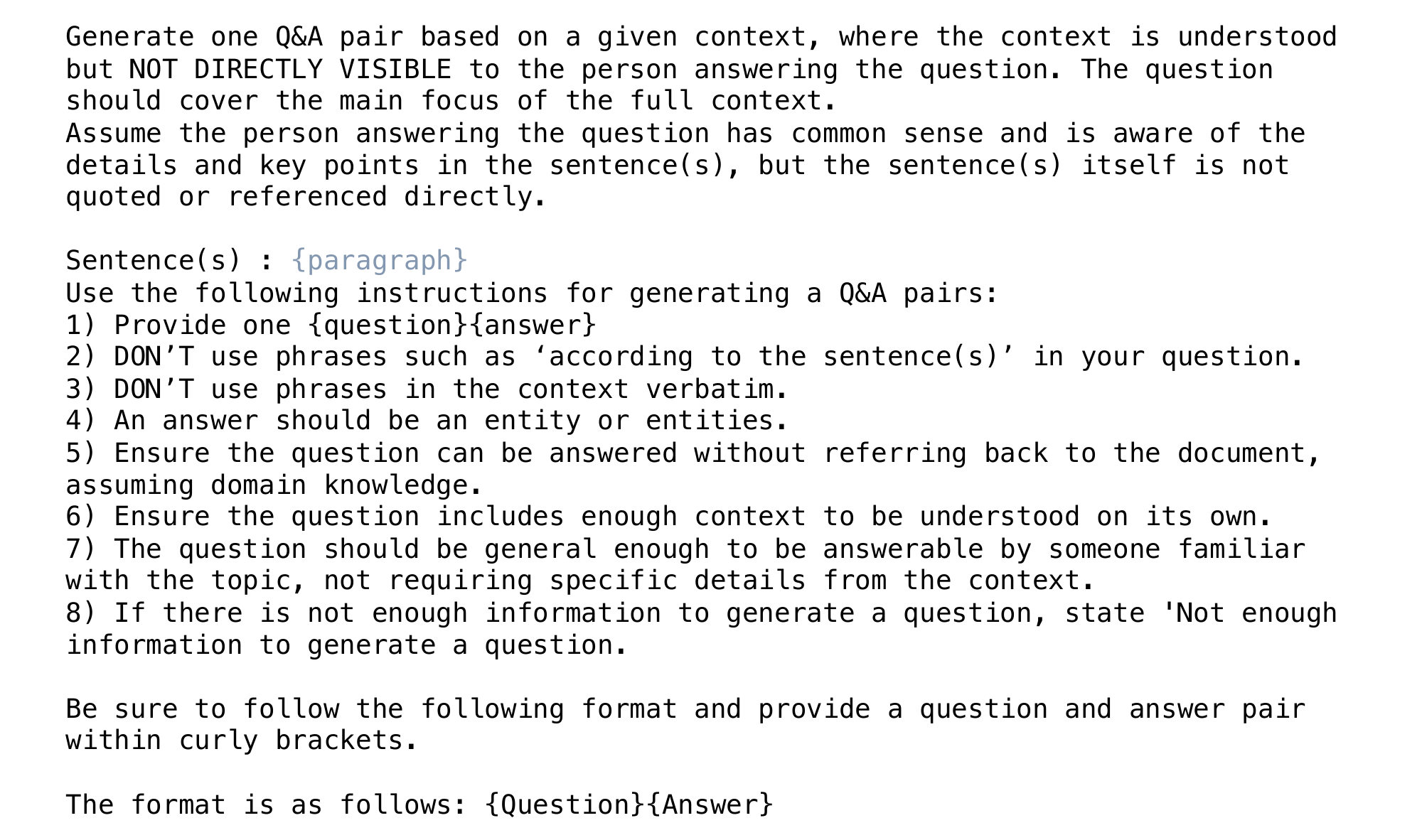}
    \caption{
    The prompt template to create pseudo queries using Llama3.1 8B in zero-shot. We prompt it to generate a question along with a corresponding answer to ensure the question can be answered. We use only generated question for evaluation.
    }
    \label{fig:prompt}
    \vspace{-0.4cm}
\end{figure*}

\textbf{GenQuery.} We use Llama3.1 8B with Q4\_0 quantization to generate pseudo queries with the temperature set to 0.5 and the max\_new\_tokens set to 256. Also, the prompt template is shown in Figure~\ref{fig:prompt}.

\textbf{IPQ.} For production quantization, we set the number of quantization groups as 8, which means the last hidden state after pooling (e.g., 1024 dimensions) is divided into 8 groups (e.g., 128 dimensions for each group), and each is clustered using KMeans. We set the number of clusters as 16, which means each 128-dimensional vector becomes an integer between 0 and 15.  

\textbf{Continual Updates.} For the training dataset, we use generated queries from the Hugging Face BEIR repository to retrain the retriever, as original test queries in the BEIR are used in the evaluation. Using these queries, we perform supervised fine-tuning. We set epoch to 4, gradient\_accumulation\_steps to 256, batch\_size to 4, learning\_rate to 1e-04, lr\_scheduler to "Constant", with multi-GPU (4 GPUs) parallelization.

\section{Concerns Regarding Computational Cost and Large Corpora}

During the rebuttal period, anonymous reviewers raised insightful and valid concerns regarding computational efficiency and scalability. Here we address these concerns in detail.

\subsection{Computational Cost}

\paragraph{Gradient computation concern.} We emphasize that our GradNormIR method remains practical even for large-scale retriever models exceeding one billion parameters. Specifically, we successfully applied GradNormIR to the GTR-XL retriever model (1.24B parameters) \citep{ni2021largedualencodersgeneralizable} and confirmed gradient norm computations were feasible on a single GPU with 48GB of memory. Furthermore, parallelizing computations across four GPUs allowed us to process the entire Arguana corpus (approximately 1.4K documents) in roughly 20 minutes. 

It is important to clarify that the computational complexity of GradNormIR primarily depends on the number of negative samples utilized in the InfoNCE loss computation. Crucially, we observed effective GradNorm computations with as few as four negative samples, significantly fewer than the typical 255 negatives per positive used during standard dense retriever training (e.g., Contriever training).

\paragraph{Runtime Cost Analysis.} We conducted a detailed runtime analysis of GradNormIR across different retriever models and varying corpus sizes using four GPUs (24GB each) in parallel. The runtime costs are summarized in Table~\ref{tab:runtime_cost}.

\begin{table*}[t!]
\centering
\resizebox{\linewidth}{!}{
\begin{tabular}{lcccc}
\toprule
\textbf{Retriever (Size)} & \textbf{Task} & \textbf{Arguana (1.4K docs)} & \textbf{Scidocs (4K docs)} & \textbf{COVID (15K docs)} \\
\midrule
\multirow{3}{*}{Contriever (110M)} & Retrieval & 2m & 3m & 6m \\
 & Gradient norm computation & 8m & 40m & 2h 30m \\
 & Total & 10m & 43m & 2h 45m \\
\midrule
\multirow{3}{*}{E5 (560M)} & Retrieval & 3m & 5m & 25m \\
 & Gradient norm computation & 12m & 1h & 3h 30m \\
 & Total & 15m & 1h 5m & 3h 55m \\
\bottomrule
\end{tabular}}
\caption{Runtime analysis for GradNormIR on different retrievers and corpus sizes.}
\label{tab:runtime_cost}
\end{table*}

For large corpora, computational costs can be effectively reduced by randomly sampling documents or partitioning the corpus. To validate this, we conducted additional experiments on large-scale datasets (DBPedia, FiQA, Quora, and COVID) by randomly sampling only 10\% of the documents. The results, shown in Table~\ref{tab:sampling_effect}, indicate minimal differences in document retrieval rates (DRR), confirming the practicality and scalability of GradNormIR through subset sampling.

\begin{table}[t!]
\centering
\resizebox{\linewidth}{!}{
\begin{tabular}{lccccc}
\toprule
\textbf{Retriever} & \textbf{Sampling} & \textbf{DBPedia} & \textbf{FiQA} & \textbf{Quora} & \textbf{COVID} \\
\midrule
\multirow{2}{*}{BGE} & w/o Sampling & 31.49 & 79.16 & 99.71 & 15.22 \\
 & w/ Sampling & 30.33 & 79.26 & 98.57 & 15.86 \\
\midrule
\multirow{2}{*}{Contriever} & w/o Sampling & 40.63 & 56.12 & 98.75 & 8.09 \\
 & w/ Sampling & 39.89 & 52.69 & 99.41 & 9.33 \\
\midrule
\multirow{2}{*}{E5} & w/o Sampling & 29.74 & 74.47 & 99.68 & 15.85 \\
 & w/ Sampling & 29.78 & 75.75 & 99.27 & 15.54 \\
\midrule
\multirow{2}{*}{GTE} & w/o Sampling & 51.24 & 71.02 & 99.60 & 16.56 \\
 & w/ Sampling & 53.87 & 70.20 & 99.20 & 16.10 \\
\bottomrule
\end{tabular}}
\caption{Comparison of Document Retrieval Rates (DRR) with and without 10\% sampling, demonstrating minimal impact on retrieval performance.}
\label{tab:sampling_effect}
\end{table}
\section{Ablation Study}\label{app:ablation}

We conduct additional ablation study of the impact of (i) the number of documents randomly sampled from the in-domain dataset and (ii) the number of positives in Eq. (\ref{eq:gradnormir}) as well as dropout rate.

\subsection{The Number of In-Domain Documents}

Table~\ref{tab:ablation-nq-1000} shows the results of DRR in predicting OOD documents, where the number of documents are determined by randomly selected 1,000 NQ documents, while Table~\ref{tab:ablation-nq-2000} shows the results when 2,000 NQ documents are used for in-domain document samples. In both cases, our method show the lowest average DDR results for all models, indicating the robustness of GradNormIR in predicting OOD documents. Also, the number of documents detected as OOD are presented in Table~\ref{tab:ablation-num-failures}. The number of OOD documents are lowest in GTE, as it can generalize to the datasets of the BEIR benchmark. 

\subsection{The Number of Positives and Dropout Rate}\label{sec:ablation-dropout}

The DRR results for the number of positives from 1 to 16 are shown in Table~\ref{tab:ablation-pos}. As the number of positives increases, the DRR generally decreases because more gradient norm values make the method more robust. Additionally, when comparing the cases with and without dropout, the decrease is significantly higher as the number of positives increases. This is because the lower-ranking positives are more likely to be affected by dropout. However, when the dropout rate increases from 0.02 to 0.05, there are some cases where the filtered documents show higher DRR values, especially increasing 3.79 in an average of 16 samples for E5. This may be because excessive dropout can deteriorate model performance.

\begin{table*}[b!]
\begin{center}
\renewcommand{\arraystretch}{1.}
\resizebox{\linewidth}{!}{
\begin{tabular}{p{1.7cm} >{\arraybackslash}p{3cm} > {\centering\arraybackslash}p{1.3cm} >{\centering\arraybackslash}p{1.7cm} >{\centering\arraybackslash}p{1.3cm} >{\centering\arraybackslash}p{1.3cm} >{\centering\arraybackslash}p{1.3cm} >{\centering\arraybackslash}p{1.3cm} >{\centering\arraybackslash}p{1.3cm} >{\centering\arraybackslash}p{1.3cm} >{\centering\arraybackslash}p{1.3cm} >{\centering\arraybackslash}p{1.3cm}|>{\centering\arraybackslash}p{1.3cm}} 
    \toprule
    \textbf{Retriever} & \textbf{Documents} & ArguAna & C-FEVER & DBPedia & FiQA & NFCorpus & Quora & Scidocs & SciFact & COVID & Touch\'e & Avg ($\downarrow$) \\ \midrule
    \multirow{5}{*}{\textbf{BGE}} 
        & ALL & 99.68 & 79.96 & 59.67 & 80.25 & 21.39 & 99.68 & 72.33 & 99.76 & 16.53 & 98.45 & 73.48 \\ \cmidrule{2-13} 
        & OOD w/ GenQuery & 100.0 & 86.87 & 75.48 & 79.7 & 21.42 & \textbf{98.96} & 62.13 & 100.0 & 15.97 & 97.4 & 73.79 \\
        & OOD w/ Layerwise & \textbf{99.01} & 86.14 & 45.48 & 79.73 & 22.35 & 99.78 & 61.95 & 100.0 & \textbf{15.34} & 93.33 & 70.31 \\
        & OOD w/ IPQ & 100.0 & 74.65 & 55.02 & 81.75 & 19.11 & 100.0 & 82.24 & \textbf{99.72} & 15.6 & 100.0 & 72.81 \\
        & OOD w/ Ours & 99.08 & \textbf{63.9} & \textbf{32.57} & \textbf{79.34} & \textbf{18.78} & 99.74 & \textbf{58.3} & 100.0 & 15.43 & \textbf{89.87} & \textbf{65.7} \\ \midrule
    \multirow{5}{*}{\textbf{Contriever}} 
        & ALL & 96.79 & 72.40 & 56.76 & 59.83 & 18.66 & 98.83 & 55.26 & 98.25 & 9.14 & 96.14 & 66.06 \\ \cmidrule{2-13} 
        & OOD w/ GenQuery & \textbf{90.12} & 72.23 & 65.87 & \textbf{54.99} & 18.53 & \textbf{97.28} & 51.17 & 98.65 & \textbf{7.45} & 93.33 & 64.96 \\
        & OOD w/ Layerwise & 93.83 & 68.67 & 49.14 & 56.61 & 17.85 & 99.1 & 51.75 & 98.36 & 8.26 & 95.34 & 63.89 \\
        & OOD w/ IPQ & 93.83 & 69.11 & 48.37 & 57.72 & 18.11 & 99.17 & 51.39 & 98.7 & 8.25 & 94.04 & 63.87 \\
        & OOD w/ Ours & 91.01 & \textbf{64.45} & \textbf{41.38} & 56.65 & \textbf{17.23} & 98.75 & \textbf{50.93} & \textbf{97.89} & 8.27 & \textbf{91.04} & \textbf{61.76} \\ \midrule
    \multirow{5}{*}{\textbf{E5}} 
        & ALL & 99.68 & 76.42 & 55.56 & 74.85 & 18.03 & 99.67 & 61.49 & 98.49 & 15.81 & 97.75 & 70.00\\ \cmidrule{2-13} 
        & OOD w/ GenQuery & \textbf{98.91} & 80.15 & 69.33 & 74.41 & 18.54 & \textbf{99.51} & 58.05 & 98.53 & 15.79 & 98.03 & 71.13 \\ 
        & OOD w/ Layerwise & 100.0 & 75.46 & 47.01 & \textbf{74.38} & 19.04 & 99.65 & \textbf{55.53} & \textbf{98.43} & 15.96 & 97.81 & 68.33 \\
        & OOD w/ IPQ & \textbf{98.91} & 75.96 & 49.54 & 74.61 & 18.01 & 99.83 & 58.57 & 98.45 & \textbf{15.68} & 97.24 & 68.68 \\
        & OOD w/ Ours & 99.48 & \textbf{69.46} & \textbf{30.27} & 74.53 & \textbf{17.14} & 99.66 & 55.79 & 98.59 & 15.84 &\textbf{96.2} & \textbf{65.7} \\ \midrule
    \multirow{5}{*}{\textbf{GTE}} 
        & ALL & 99.68 & 80.37 & 60.85 & 75.76 & 22.48 & 99.57 & 72.66 & 99.52 & 17.53 & 99.25 & 73.55\\ \cmidrule{2-13} 
        & OOD w/ GenQuery & 100.0 & 84.2 & 76.8 & \textbf{70.58} & 21.33 & \textbf{98.71} & 65.95 & 99.73 & 16.16 & 99.63 & 73.31 \\
        & OOD w/ Layerwise & 100.0 & 82.81 & 56.66 & 76.17 & 21.48 & 99.87 & 66.14 & \textbf{99.47} & 14.98 & 99.82 & 71.74 \\
        & OOD w/ IPQ & 100.0 & 84.59 & 66.26 & 75.27 & 19.82 & 99.83 & 68.45 & 99.5 & \textbf{14.81} & 99.82 & 72.84 \\
        & OOD w/ Ours & \textbf{93.75} & \textbf{70.73} & \textbf{51.49} & 70.97 & \textbf{19.27} & 99.59 & \textbf{65.25} & 100.0 & 16.52 & \textbf{98.72} & \textbf{68.63} \\
    \bottomrule
\end{tabular}
}
\caption{
    Comparison of OOD document detection across  different retriever models, with the number of documents selected by \textbf{1,000} sampled NQ documents.
}
\label{tab:ablation-nq-1000}
\end{center}
\end{table*}

\begin{table*}[b!]
\begin{center}
\renewcommand{\arraystretch}{1.}
\resizebox{\linewidth}{!}{
\begin{tabular}{p{1.7cm} >{\arraybackslash}p{3cm} > {\centering\arraybackslash}p{1.3cm} >{\centering\arraybackslash}p{1.7cm} >{\centering\arraybackslash}p{1.3cm} >{\centering\arraybackslash}p{1.3cm} >{\centering\arraybackslash}p{1.3cm} >{\centering\arraybackslash}p{1.3cm} >{\centering\arraybackslash}p{1.3cm} >{\centering\arraybackslash}p{1.3cm} >{\centering\arraybackslash}p{1.3cm} >{\centering\arraybackslash}p{1.3cm}|>{\centering\arraybackslash}p{1.3cm}} 
    \toprule
    \textbf{Retriever} & \textbf{Documents} & ArguAna & C-FEVER & DBPedia & FiQA & NFCorpus & Quora & Scidocs & SciFact & COVID & Touch\'e & Avg ($\downarrow$) \\ \midrule
    \multirow{5}{*}{\textbf{BGE}} 
        & ALL & 99.68 & 79.96 & 59.67 & 80.25 & 21.39 & 99.68 & 72.33 & 99.76 & 16.53 & 98.45 & 73.48 \\ \cmidrule{2-13} 
        & OOD w/ GenQuery & 100.0 & 86.87 & 75.48 & 79.7 & 21.42 & \textbf{98.96} & 62.13 & 100.0 & 15.97 & 97.4 & 73.79 \\
        & OOD w/ Layerwise & \textbf{99.01} & 86.14 & 45.48 & 79.73 & 22.35 & 99.78 & 61.95 & 100.0 & \textbf{15.34} & 93.33 & 70.31 \\
        & OOD w/ IPQ & 100.0 & 74.65 & 55.02 & 81.75 & 19.11 & 100.0 & 82.24 & \textbf{99.72} & 15.6 & 100.0 & 72.81 \\
        & OOD w/ Ours & 99.03 & \textbf{64.18} & \textbf{31.97} & \textbf{79.27} & \textbf{18.42} & 99.73 & \textbf{57.25} & 100.0 & 15.35 & \textbf{89.47} & \textbf{65.47} \\ \midrule
    \multirow{5}{*}{\textbf{Contriever}} 
        & ALL & 96.79 & 72.40 & 56.76 & 59.83 & 18.66 & 98.83 & 55.26 & 98.25 & 9.14 & 96.14 & 66.06 \\ \cmidrule{2-13} 
        & OOD w/ GenQuery & \textbf{90.12} & 72.23 & 65.87 & \textbf{54.99} & 18.53 & \textbf{97.28} & 51.17 & 98.65 & \textbf{7.45} & 93.33 & 64.96 \\
        & OOD w/ Layerwise & 93.83 & 68.67 & 49.14 & 56.61 & 17.85 & 99.1 & 51.75 & 98.36 & 8.26 & 95.34 & 63.89 \\
        & OOD w/ IPQ & 93.83 & 69.11 & 48.37 & 57.72 & 18.11 & 99.17 & 51.39 & 98.7 & 8.25 & 94.04 & 63.87 \\
        & OOD w/ Ours & 91.76 & \textbf{64.23} & \textbf{40.88} & 56.37 & \textbf{17.17} & 98.73 & \textbf{50.58} & \textbf{97.79} & 8.17 & \textbf{89.88} & \textbf{61.56} \\ \midrule
    \multirow{5}{*}{\textbf{E5}} 
        & ALL & 99.68 & 76.42 & 55.56 & 74.85 & 18.03 & 99.67 & 61.49 & 98.49 & 15.81 & 97.75 & 70.00\\ \cmidrule{2-13} 
        & OOD w/ GenQuery & \textbf{98.91} & 80.15 & 69.33 & 74.41 & 18.54 & \textbf{99.51} & 58.05 & 98.53 & 15.79 & 98.03 & 71.13 \\ 
        & OOD w/ Layerwise & 100.0 & 75.46 & 47.01 & \textbf{74.38} & 19.04 & 99.65 & \textbf{55.53} & \textbf{98.43} & 15.96 & 97.81 & 68.33 \\
        & OOD w/ IPQ & \textbf{98.91} & 75.96 & 49.54 & 74.61 & 18.01 & 99.83 & 58.57 & 98.45 & \textbf{15.68} & 97.24 & 68.68 \\
        & OOD w/ Ours & 99.47 & \textbf{69.1} & \textbf{29.97} & 74.47 & \textbf{17.09} & 99.67 & 55.82 & 98.56 & 15.85 & \textbf{96.48} & \textbf{65.65} \\ \midrule
    \multirow{5}{*}{\textbf{GTE}} 
        & ALL & 99.68 & 80.37 & 60.85 & 75.76 & 22.48 & 99.57 & 72.66 & 99.52 & 17.53 & 99.25 & 73.55\\ \cmidrule{2-13} 
        & OOD w/ GenQuery & 100.0 & 84.2 & 76.8 & \textbf{70.58} & 21.33 & 98.71 & 65.95 & 99.73 & 16.16 & 99.63 & 73.31 \\
        & OOD w/ Layerwise & 100.0 & 82.81 & 56.66 & 76.17 & 21.48 & 99.87 & 66.14 & \textbf{99.47} & 14.98 & 99.82 & 71.74\\
        & OOD w/ IPQ & 100.0 & 84.59 & 66.26 & 75.27 & 19.82 & 99.83 & 68.45 & 99.5 & \textbf{14.81} & 99.82 & 72.84\\
        & OOD w/ Ours & \textbf{93.75} & \textbf{70.71} & \textbf{51.17} & 71.02 & \textbf{19.21} & \textbf{99.6} & \textbf{65.25} & 100.0 & 16.56 & \textbf{98.72} & \textbf{68.6} \\
    \bottomrule
\end{tabular}
}
\caption{
     Comparison of OOD document detection across  different retriever models, with the number of documents selected by \textbf{2,000} sampled NQ documents.
}
\label{tab:ablation-nq-2000}
\end{center}
\end{table*}

\begin{table*}[b!]
\begin{center}
\renewcommand{\arraystretch}{1}
\resizebox{\linewidth}{!}{
\begin{tabular}{p{1.5cm} >{\centering\arraybackslash}p{1.9cm}|>{\centering\arraybackslash}p{1.3cm} >{\centering\arraybackslash}p{1.3cm} >{\centering\arraybackslash}p{1.3cm} >{\centering\arraybackslash}p{1.3cm} >{\centering\arraybackslash}p{1.3cm} >{\centering\arraybackslash}p{1.3cm} >{\centering\arraybackslash}p{1.3cm} >{\centering\arraybackslash}p{1.3cm} >{\centering\arraybackslash}p{1.3cm} >{\centering\arraybackslash}p{1.3cm}} 
    \toprule
    \textbf{Retriever} & \textbf{\# Samples} & ArguAna & Climate-FEVER & DBPedia & FiQA & NFCorpus & Quora & Scidocs & SciFact & TREC-COVID & Touch\'e \\ \midrule
    \multirow{3}{*}{BGE}
        & 1000 & 109 & 339 & 6940 & 15686 & 1236 & 1542 & 1204 & 221 & 10018 & 79 \\
        & 2000 & 103 & 323 & 6795 & 15593 & 1143 & 1458 & 1139 & 207 & 9632 & 76 \\
        & 3000 & 101 & 306 & 6660 & 15489 & 1063 & 1376 & 1073 & 190 & 9194 & 74 \\
    \midrule
    \multirow{3}{*}{\textbf{Contriever}} 
        & 1000 & 89 & 571 & 9712 & 13201 & 2429 & 6320 & 2926 & 505 & 10469 & 278 \\
        & 2000 & 85 & 540 & 9492 & 12655 & 2328 & 5922 & 2830 & 483 & 10005 & 246 \\
        & 3000 & 81 & 520 & 9378 & 12343 & 2271 & 5695 & 2769 & 480 & 9742 & 233 \\
    \midrule
    \multirow{3}{*}{\textbf{E5}} 
        & 1000 & 192 & 855 & 6035 & 16383 & 2379 & 7347 & 2591 & 523 & 14496 & 520 \\
        & 2000 & 187 & 836 & 5850 & 16328 & 2319 & 7059 & 2534 & 508 & 14464 & 505 \\
        & 3000 & 183 & 815 & 5736 & 16297 & 2289 & 6895 & 2500 & 506 & 14437 & 497 \\
    \midrule
    \multirow{3}{*}{\textbf{GTE}} 
        & 1000 & 16 & 494 & 3643 & 7095 & 469 & 2931 & 1828 & 193 & 7977 & 536 \\
        & 2000 & 16 & 499 & 3759 & 7259 & 486 & 2990 & 1853 & 201 & 8113 & 537 \\
        & 3000 & 16 & 497 & 3738 & 7219 & 478 & 2973 & 1846 & 197 & 8082 & 537 \\
    \bottomrule
\end{tabular}
}
\caption{
    The number of detected OOD documents for each dataset, determined by the randomly sampled 1000, 2000, and 3000 NQ documents.
}
\label{tab:ablation-num-failures}
\end{center}
\end{table*}

\begin{table*}[b!]
\begin{center}
\renewcommand{\arraystretch}{1}
\resizebox{\linewidth}{!}{
\begin{tabular}{p{1.9cm} >{\centering\arraybackslash}p{1.3cm}>{\centering\arraybackslash}p{1.3cm} >{\centering\arraybackslash}p{1.3cm} >{\centering\arraybackslash}p{1.3cm} >{\centering\arraybackslash}p{1.3cm} >{\centering\arraybackslash}p{1.3cm} >{\centering\arraybackslash}p{1.3cm} >{\centering\arraybackslash}p{1.3cm} >{\centering\arraybackslash}p{1.3cm} >{\centering\arraybackslash}p{1.3cm} >{\centering\arraybackslash}p{1.3cm}} 
    \toprule
    \textbf{Retriever} & ArguAna & Climate-FEVER & DBPedia & FiQA & NFCorpus & Quora & Scidocs & SciFact & TREC-COVID & Touch\'e \\ \midrule
    \textbf{BGE} & 96.83 & 24.42 & 3.65 & 48.4 & 5.45 & 99.59 & 14.04 & 100.0 & 1.94 & 9.38 \\
    \midrule
    \textbf{Contriever} & 81.97 & 30.43 & 8.0 & 26.53 & 4.93 & 96.06 & 23.27 & 89.81 & 0.76 & 23.53 \\
    \midrule
    \textbf{E5} & 95.07 & 25.12 & 4.67 & 43.45 & 4.97 & 98.72 & 21.78 & 93.69 & 1.9 & 24.61 \\
    \midrule
    \textbf{GTE} & 100.0 & 39.81 & 30.56 & 42.88 & 7.81 & 91.13 & 26.87 & 81.82 & 2.48 & 40.53 \\
    \bottomrule
\end{tabular}
}
\caption{
    DRR with Recall@10.
}
\end{center}
\end{table*}

\begin{table*}[b!]
\begin{center}
\renewcommand{\arraystretch}{1}
\resizebox{\linewidth}{!}{
\begin{tabular}{p{1.9cm} >{\centering\arraybackslash}p{1.3cm}>{\centering\arraybackslash}p{1.3cm} >{\centering\arraybackslash}p{1.3cm} >{\centering\arraybackslash}p{1.3cm} >{\centering\arraybackslash}p{1.3cm} >{\centering\arraybackslash}p{1.3cm} >{\centering\arraybackslash}p{1.3cm} >{\centering\arraybackslash}p{1.3cm} >{\centering\arraybackslash}p{1.3cm} >{\centering\arraybackslash}p{1.3cm}} 
    \toprule
    \textbf{Retriever} & ArguAna & Climate-FEVER & DBPedia & FiQA & NFCorpus & Quora & Scidocs & SciFact & TREC-COVID & Touch\'e \\ \midrule
    \textbf{BGE} & 98.41 & 32.26 & 9.23 & 62.69 & 8.53 & 99.79 & 23.97 & 100.0 & 4.20 & 46.88 \\
    \midrule
    \textbf{Contriever} & 91.80 & 44.31 & 18.36 & 38.19 & 8.88 & 97.47 & 34.08 & 93.57 & 2.34 & 67.65 \\
    \midrule
    \textbf{E5} & 98.03 & 34.74 & 8.38 & 57.64 & 8.74 & 99.21 & 32.46 & 97.6 & 5.44 & 69.11 \\
    \midrule
    \textbf{GTE} & 100.0 & 54.37 & 56.97 & 56.17 & 14.51 & 95.91 & 45.26 & 90.91 & 5.39 & 88.17 \\
    \bottomrule
\end{tabular}
}
\caption{
    DRR with Recall@30.
}
\end{center}
\end{table*}

\begin{table*}[b!]
\begin{center}
\renewcommand{\arraystretch}{1}
\resizebox{\linewidth}{!}{
\begin{tabular}{p{1.5cm} >{\centering\arraybackslash}p{1.9cm}|>{\centering\arraybackslash}p{1.3cm} >{\centering\arraybackslash}p{1.3cm} >{\centering\arraybackslash}p{1.3cm} >{\centering\arraybackslash}p{1.3cm} >{\centering\arraybackslash}p{1.3cm} >{\centering\arraybackslash}p{1.3cm} >{\centering\arraybackslash}p{1.3cm} >{\centering\arraybackslash}p{1.3cm} >{\centering\arraybackslash}p{1.3cm} >{\centering\arraybackslash}p{1.3cm} >{\centering\arraybackslash}p{1.3cm}} 
    \toprule
    \textbf{Retriever} & \textbf{\# Samples} & ArguAna & Climate-FEVER & DBPedia & FiQA & NFCorpus & Quora & Scidocs & SciFact & TREC-COVID & Touch\'e \\ \midrule
    \multirow{3}{*}{\textbf{BGE}}
        & 1000 & 8.71 & 10.12 & 25.63 & 68.71 & 5.42 & 2.0 & 7.39 & 6.0 & 14.4 & 3.15 \\
        & 2000 & 8.71 & 10.19 & 25.65 & 68.82 & 5.48 & 2.01 & 7.41 & 6.0 & 14.49 & 3.15 \\
        & 3000 & 8.99 & 10.42 & 26.14 & 69.96 & 5.67 & 2.09 & 7.81 & 6.45 & 15.6 & 3.48\\
    \midrule
    \multirow{3}{*}{\textbf{Contriever}} 
        & 1000 & 8.21 & 22.25 & 46.57 & 41.71 & 38.32 & 13.21 & 46.89 & 49.33 & 34.88 & 10.33 \\
        & 2000 & 8.42 & 22.84 & 46.85 & 42.6 & 39.06 & 13.52 & 47.56 & 50.07 & 35.54 & 10.65 \\
        & 3000 & 8.71 & 22.84 & 47.09 & 43.25 & 39.61 & 13.74 & 47.99 & 50.22 & 36.35 & 11.09 \\
    \midrule
    \multirow{3}{*}{\textbf{E5}} 
        & 1000 & 14.49 & 19.12 & 10.67 & 79.88 & 29.51 & 8.75 & 25.82 & 27.74 & 84.29 & 20.65 \\
        & 2000 & 14.20 & 18.68 & 10.25 & 79.17 & 28.57 & 8.48 & 25.32 & 26.84 & 83.70 & 19.57 \\
        & 3000 & 14.49 & 19.12 & 10.95 & 79.85 & 29.48 & 8.73 & 25.8 & 27.74 & 84.24 & 20.65 \\
    \midrule
    \multirow{3}{*}{\textbf{GTE}} 
        & 1000 & 0.21 & 9.45 & 4.49 & 11.96 & 2.2 & 3.91 & 13.73 & 1.5 & 8.49 & 40.43 \\
        & 2000 & 0.21 & 8.85 & 4.39 & 11.63 & 2.15 & 3.78 & 13.23 & 1.35 & 7.88 & 39.78 \\
        & 3000 & 0.07 & 6.62 & 3.69 & 9.89 & 1.49 & 3.15 & 10.92 & 0.6 & 5.86 & 35.87 \\
    \bottomrule
\end{tabular}
}
\caption{
    Ratio of detected OOD documents for each dataset over total documents, determined by the randomly sampled 1000, 2000, and 3000 NQ documents.
}
\end{center}
\end{table*}

\begin{table*}[b!]
\begin{center}
\renewcommand{\arraystretch}{1}
\resizebox{\linewidth}{!}{
\begin{tabular}{p{1.8cm}>{\centering\arraybackslash}p{1.7cm} >{\centering\arraybackslash}p{1.7cm}|>{\centering\arraybackslash}p{1.3cm} >{\centering\arraybackslash}p{1.3cm} >{\centering\arraybackslash}p{1.3cm} >{\centering\arraybackslash}p{1.3cm} >{\centering\arraybackslash}p{1.3cm} >{\centering\arraybackslash}p{1.3cm} >{\centering\arraybackslash}p{1.3cm} >{\centering\arraybackslash}p{1.3cm} >{\centering\arraybackslash}p{1.3cm} >{\centering\arraybackslash}p{1.3cm}|>{\centering\arraybackslash}p{1.3cm}} 
    \toprule
    \textbf{Retriever} & \textbf{Dropout} & \textbf{Num Pos} & ArguAna & Climate-FEVER & DBPedia & FiQA & NFCorpus & Quora & Scidocs & SciFact & TREC-COVID & Touch\'e & Avg ($\downarrow$) \\ \midrule
    \multirow{15}{*}{\textbf{BGE}} 
        & \multirow{5}{*}{\xmark} 
            & 1 & 98.99 & 68.32 & 35.25 & 79.48 & 19.07 & 99.85 & 62.51 & 100.0 & 15.06 & 89.19 & 66.77 \\
            & & 2 & 98.99 & 62.57 & 33.65 & 79.24 & 18.59 & 99.71 & 61.25 & 99.7 & 15.33 & 91.89 & 66.09 \\
            & & 4 & 98.99 & 60.23 & 32.66 & 79.21 & 18.43 & 99.71 & 60.96 & 99.72 & 14.89 & 89.19 & 65.4 \\
            & & 8 & 98.99 & 62.34 & 31.53 & 79.13 & 18.64 & 99.78 & 61.06 & 99.72 & 14.83 & 91.89 & 65.79 \\
            & & 16 & 98.99 & 62.34 & 31.53 & 79.13 & 18.64 & 99.78 & 61.06 & 99.72 & 14.83 & 91.89 & 65.79 \\ \cmidrule{2-14}
        & \multirow{5}{*}{0.02} 
            & 1 & 99.01 & 67.31 & 36.78 & 79.29 & 19.37 & 99.35 & 59.1 & 100.0 & 16.17 & 89.33 & 66.57 \\
            & & 2 & 99.01 & 77.89 & 34.66 & 79.23 & 19.01 & 99.42 & 57.86 & 100.0 & 15.84 & 88.16 & 67.11 \\
            & & 4 & 99.01 & 78.24 & 33.15 & 79.22 & 18.72 & 99.49 & 57.77 & 100.0 & 15.74 & 90.54 & 67.19 \\
            & & 8 & 99.01 & 61.14 & 31.49 & 79.16 & 18.36 & 99.71 & 56.97 & 100.0 & 15.22 & 89.19 & 65.03 \\
            & & 16 & 99.01 & 60.65 & 31.57 & 79.15 & 18.39 & 99.85 & 60.31 & 100.0 & 15.0 & 89.19 & 65.31 \\ \cmidrule{2-14}
        & \multirow{5}{*}{0.05} 
            & 1 & 99.01 & 61.06 & 35.71 & 79.45 & 19.35 & 99.78 & 63.91 & 100.0 & 15.47 & 90.54 & 66.43 \\
            & & 2 & 99.01 & 60.0 & 33.87 & 79.37 & 18.92 & 99.78 & 62.16 & 100.0 & 15.18 & 90.54 & 65.88 \\
            & & 4 & 99.01 & 59.27 & 32.42 & 79.28 & 18.42 & 99.85 & 61.74 & 100.0 & 15.06 & 89.19 & 65.42 \\
            & & 8 & 99.01 & 58.76 & 31.63 & 79.15 & 18.21 & 99.85 & 61.52 & 100.0 & 15.02 & 89.19 & 65.24 \\
            & & 16 & 99.01 & 58.76 & 31.63 & 79.15 & 18.21 & 99.85 & 61.52 & 100.0 & 15.02 & 89.19 & 65.24 \\ \midrule

    \multirow{15}{*}{\textbf{Contriever}} 
        & \multirow{5}{*}{\xmark} 
            & 1 & 92.41 & 71.08 & 45.11 & 57.55 & 17.54 & 99.32 & 52.87 & 97.79 & 8.25 & 92.7 & 63.36 \\
            & & 2 & 92.41 & 70.45 & 44.47 & 57.25 & 17.63 & 99.24 & 52.58 & 98.02 & 8.04 & 91.85 & 63.09 \\
            & & 4 & 90.12 & 71.08 & 45.11 & 57.55 & 17.54 & 99.32 & 52.87 & 97.79 & 8.25 & 92.7 & 63.36 \\
            & & 8 & 91.36 & 64.35 & 43.68 & 57.07 & 17.42 & 99.42 & 51.72 & 97.79 & 7.77 & 93.56 & 62.41 \\
            & & 16 & 92.59 & 64.35 & 43.68 & 57.07 & 17.42 & 99.42 & 51.72 & 97.79 & 7.77 & 93.56 & 62.54 \\ \cmidrule{2-14}
        & \multirow{5}{*}{0.02} 
            & 1 & 92.59 & 71.19 & 43.18 & 56.32 & 17.39 & 97.91 & 50.97 & 97.8 & 8.6 & 91.88 & 62.78 \\
            & & 2 & 92.59 & 69.85 & 41.87 & 56.18 & 17.09 & 98.14 & 50.01 & 97.83 & 8.44 & 91.03 & 62.30 \\
            & & 4 & 90.12 & 68.72 & 40.97 & 56.1 & 16.99 & 98.33 & 50.78 & 97.8 & 8.35 & 90.17 & 61.83 \\
            & & 8 & 91.36 & 63.92 & 40.63 & 56.12 & 17.11 & 98.75 & 50.64 & 97.78 & 8.09 & 90.17 & 61.46 \\
            & & 16 & 92.59 & 64.12 & 43.6 & 57.12 & 17.39 & 99.42 & 51.68 & 97.78 & 7.69 & 94.02 & 62.54 \\ \cmidrule{2-14}
        & \multirow{5}{*}{0.05} 
            & 1 & 92.59 & 68.93 & 41.86 & 56.19 & 17.11 & 98.16 & 49.99 & 97.83 & 8.48 & 91.03 & 62.75 \\
            & & 2 & 92.59 & 68.93 & 41.86 & 56.19 & 17.11 & 98.16 & 49.99 & 97.83 & 8.48 & 91.03 & 62.22 \\
            & & 4 & 90.12 & 68.83 & 40.99 & 56.03 & 17.01 & 98.3 & 50.92 & 97.8 & 8.28 & 90.17 & 61.85 \\
            & & 8 & 91.36 & 63.65 & 40.6 & 56.03 & 17.1 & 98.79 & 50.58 & 97.76 & 8.04 & 89.74 & 61.36 \\
            & & 16 & 92.59 & 63.85 & 43.66 & 56.88 & 17.22 & 99.42 & 51.66 & 97.79 & 7.8 & 94.42 & 62.53 \\ \midrule

    \multirow{15}{*}{\textbf{E5}} 
        & \multirow{5}{*}{\xmark} 
            & 1 & 99.45 & 75.21 & 32.43 & 74.54 & 16.91 & 99.38 & 55.11 & 98.35 & 15.83 & 97.03 & 66.42 \\
            & & 2 & 99.45 & 76.79 & 30.44 & 74.46 & 16.9 & 99.49 & 54.97 & 98.65 & 15.73 & 96.63 & 66.35 \\
            & & 4 & 99.45 & 75.25 & 29.84 & 74.5 & 16.99 & 99.61 & 55.02 & 98.75 & 15.84 & 96.25 & 66.15 \\
            & & 8 & 99.45 & 69.19 & 29.65 & 74.47 & 17.03 & 99.68 & 55.52 & 98.55 & 15.85 & 96.43 & 65.58 \\
            & & 16 & 99.45 & 66.97 & 31.21 & 74.52 & 17.05 & 99.75 & 55.72 & 98.35 & 15.71 & 96.6 & 65.53 \\ \cmidrule{2-14}
        & \multirow{5}{*}{0.02} 
            & 1 & 99.45 & 75.21 & 32.47 & 74.54 & 16.91 & 99.38 & 55.11 & 98.35 & 15.83 & 97.03 & 66.43 \\
            & & 2 & 99.45 & 76.8 & 30.44 & 74.46 & 16.9 & 99.49 & 55.01 & 98.65 & 15.73 & 96.63 & 66.36 \\
            & & 4 & 99.45 & 75.25 & 29.9 & 74.5 & 16.99 & 99.59 & 55.0 & 98.75 & 15.84 & 96.25 & 66.15\\
            & & 8 & 99.45 & 69.19 & 29.74 & 74.47 & 17.02 & 99.68 & 55.5 & 98.56 & 15.85 & 96.43 & 65.59 \\
            & & 16 & 99.45 & 66.99 & 30.99 & 74.52 & 17.09 & 99.77 & 55.71 & 98.35 & 15.72 & 96.6 & 65.52 \\ \cmidrule{2-14}
        & \multirow{5}{*}{0.05} 
            & 1 & 99.51 & 64.88 & 40.74 & 74.41 & 16.79 & 99.8 & 53.42 & 98.83 & 14.95 & 97.4 & 66.07 \\
            & & 2 & 99.51 & 62.81 & 40.92 & 74.69 & 16.59 & 99.85 & 52.99 & 98.21 & 15.04 & 97.92 & 65.85 \\
            & & 4 & 99.51 & 61.76 & 41.09 & 74.6 & 16.41 & 99.8 & 52.45 & 98.83 & 15.19 & 96.88 & 65.70 \\
            & & 8 & 99.51 & 63.96 & 42.99 & 74.8 & 15.68 & 99.9 & 51.86 & 99.12 & 15.41 & 98.96 & 66.22 \\
            & & 16 & 99.51 & 63.96 & 42.94 & 74.79 & 15.65 & 99.9 & 51.86 & 99.12 & 15.42 & 98.96 & 66.21 \\ \midrule

    \multirow{15}{*}{\textbf{GTE}} 
        & \multirow{5}{*}{\xmark} 
            & 1 & 93.75 & 80.05 & 53.77 & 73.15 & 21.77 & 99.66 & 69.18 & 99.73 & 16.31 & 98.9 & 70.63 \\
            & & 2 & 93.75 & 83.22 & 54.14 & 73.14 & 21.14 & 99.73 & 67.8 & 100.0 & 15.89 & 98.9 & 70.77 \\
            & & 4 & 93.75 & 80.18 & 57.45 & 72.14 & 22.52 & 99.66 & 67.73 & 99.72 & 15.89 & 98.9 & 70.79 \\
            & & 8 & 93.75 & 78.16 & 60.09 & 72.09 & 21.62 & 99.66 & 67.42 & 99.73 & 16.24 & 98.72 & 70.75 \\
            & & 16 & 93.75 & 78.16 & 60.09 & 72.09 & 21.62 & 99.66 & 67.42 & 99.73 & 16.24 & 98.72 & 70.75 \\ \cmidrule{2-14}
        & \multirow{5}{*}{0.02} 
            & 1 & 93.75 & 76.5 & 41.71 & 69.61 & 19.6 & 99.63 & 64.1 & 100.0 & 17.7 & 99.27 & 68.19 \\
            & & 2 & 93.75 & 82.36 & 43.65 & 69.89 & 18.5 & 99.66 & 64.2 & 100.0 & 17.32 & 99.09 & 68.84 \\
            & & 4 & 93.75 & 75.33 & 46.1 & 70.26 & 19.0 & 99.56 & 65.08 & 100.0 & 17.1 & 98.72 & 68.49 \\
            & & 8 & 93.75 & 70.83 & 51.24 & 71.02 & 19.22 & 99.6 & 65.22 & 100.0 & 16.56 & 98.72 & 68.62 \\
            & & 16 & 93.75 & 78.03 & 60.04 & 72.05 & 19.71 & 99.66 & 67.44 & 99.73 & 15.87 & 98.9 & 70.52 \\ \cmidrule{2-14}
        & \multirow{5}{*}{0.05} 
            & 1 & 93.75 & 76.54 & 41.97 & 69.7 & 19.59 & 99.56 & 64.2 & 100.0 & 17.67 & 99.27 & 68.23 \\
            & & 2 & 93.75 & 82.41 & 43.45 & 69.89 & 18.44 & 99.66 & 64.26 & 100.0 & 17.33 & 99.09 & 68.83 \\
            & & 4 & 93.75 & 75.31 & 46.52 & 70.25 & 19.06 & 99.56 & 65.08 & 100.0 & 17.12 & 98.72 & 68.54 \\
            & & 8 & 93.75 & 70.5 & 51.17 & 71.09 & 19.28 & 99.6 & 65.28 & 100.0 & 16.54 & 98.72 & 68.59 \\
            & & 16 & 93.75 & 78.41 & 60.08 & 72.12 & 19.61 & 99.66 & 67.37 & 99.73 & 16.01 & 98.9 & 70.56 \\ 
    \bottomrule
\end{tabular}
}
\caption{
    Ablation study on the number of positives for computing GradNormIR in ($p$ in Eq. (~\ref{eq:gradnormir})). The results present the DRR values of OOD documents prediction, comparing results without dropout and with dropout rates of 0.02 and 0.05 for the document query.
}
\label{tab:ablation-pos}
\end{center}
\end{table*}

\section{Relevance Gains via Filtering}

\begin{figure}[t]
\includegraphics[width=0.47\textwidth]{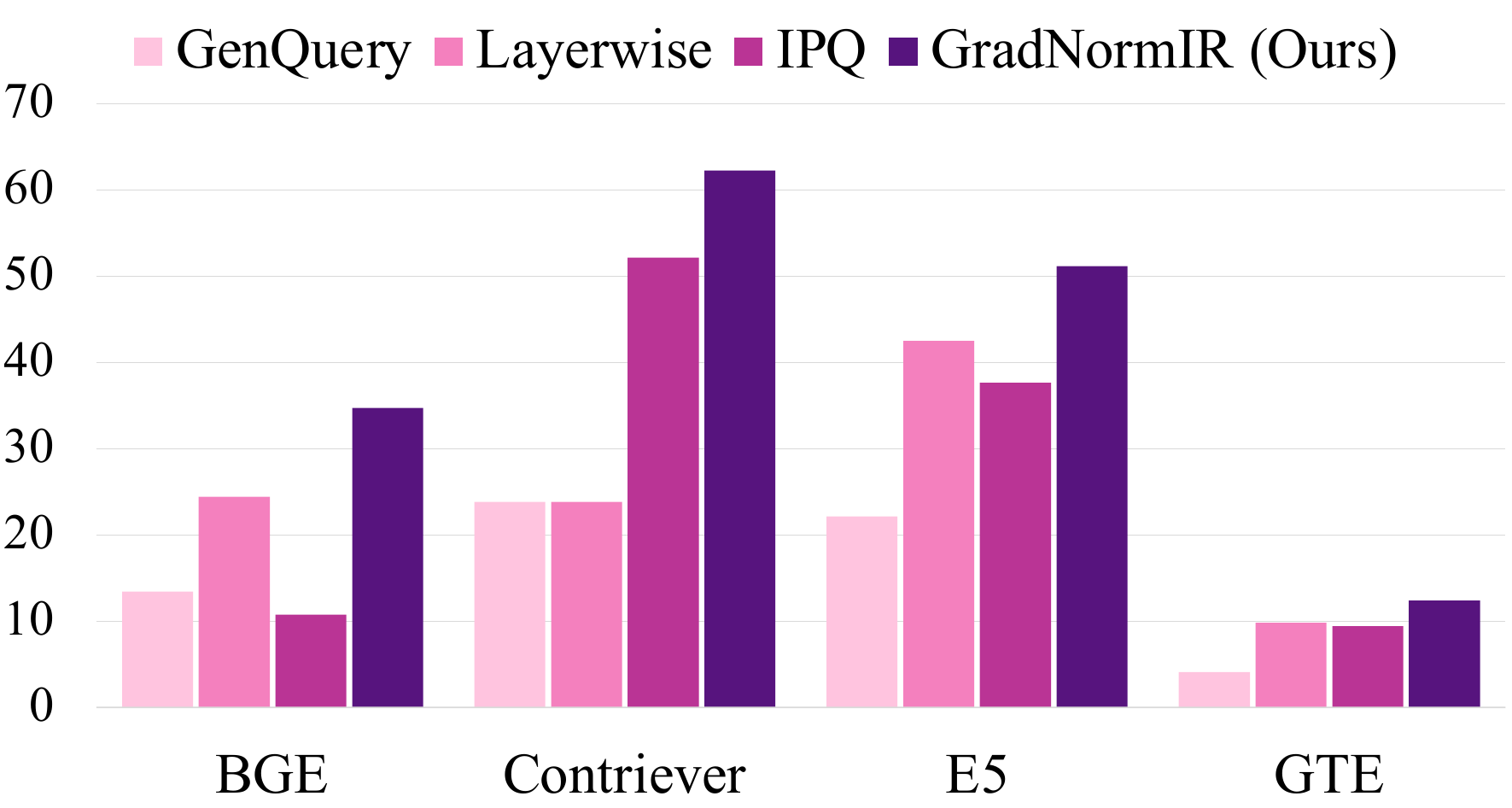}
    \caption{
    Results of relevance gains via OOD document filtering.
    }
    \label{fig:filtering}
\end{figure}

To evaluate the impact of OOD documents, we also conduct a document filtering experiment. Specifically, we remove OOD documents from the given corpus $\mathcal{C}$, thereby enhancing retrieval relevance. 

\paragraph{Setup.}
For each dataset, we begin with an evaluation set \(\{(d_i, Q_i)\}_{i=1}^{N}\). If \(d_i\) is detected as an OOD document, we remove it from the evaluation set, 
meaning we no longer evaluate \(d_i\) as a gold label for its associated queries \(q_i \in Q_i\). We then evaluate the performance on the test queries \(\{Q_i\}_{i=1}^{N}\). By removing such OOD documents, we aim to exclude irrelevant or misleading texts that could otherwise confuse the retriever, thereby potentially improving retrieval performance.
We measure retrieval performance using Recall@100, following \citet{izacardunsupervised}.

Figure~\ref{fig:filtering} presents the total sum of gains in Recall@100 across 10 datasets of the BEIR after removing OOD documents. Our method, GradNormIR, demonstrates significant performance improvements across all retrievers. Specifically, it achieves gains of 34.73, 62.24, 51.15, and 12.40 points for BGE, Contriever, E5, and GTE, respectively.
Even with GTE, where GradNormIR does not yield the best results, the overall retrieval enhancement remains the highest with our method. 

\begin{figure}[t!]
\includegraphics[width=0.5\textwidth]
{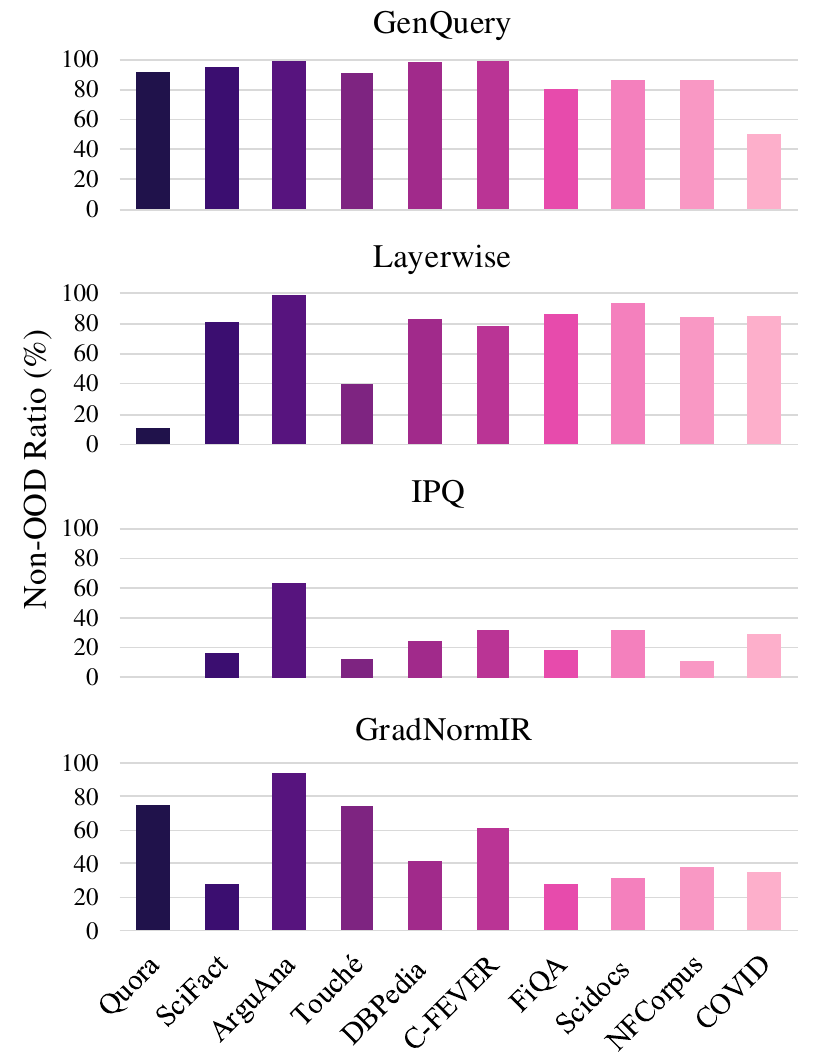}
    \caption{
    Relation Between OOD Ratio and Performance
    }
    \label{fig:ood-ratio}
\end{figure}

\section{Relation Between OOD Ratio and Performance}

Figure~\ref{fig:ood-ratio} shows the relationship between OOD document ratio $r(\mathcal{C})$ and retriever performance. The x-axis lists datasets in descending order of performance based on Contriever's \citep{izacardunsupervised} Recall@100. The y-axis represents the Non-OOD ratio (1 - $r(\mathcal{C})$). 

The graph's descending trend indicates that 1 - $r(\mathcal{C})$ is proportional to retriever performance, as datasets with higher retrieval performance show greater Non-OOD ratios. GradNormIR clearly demonstrates this relationship, showing high Non-OOD ratios for Quora, Arguana, and Touch\'e, and low ratios for FiQA, Scidocs, NFCorpus, and COVID. While GenQuery also exhibits a descending trend, it shows minimal variation from Quora to NFCorpus, making OOD corpus detection less effective. 

To predict OOD corpora, we set $\gamma$ to 0.5 based on the average performance across all datasets. With this threshold, we identify Scifact, Touch'e, DBPedia, FiQA, Scidocs, NFCorpus, and COVID as OOD corpora.  
\section{Feasibility of GradNormIR}\label{app:feasibility}

\begin{figure*}[t!]
\hbox{
    \includegraphics[width=1.0\textwidth]{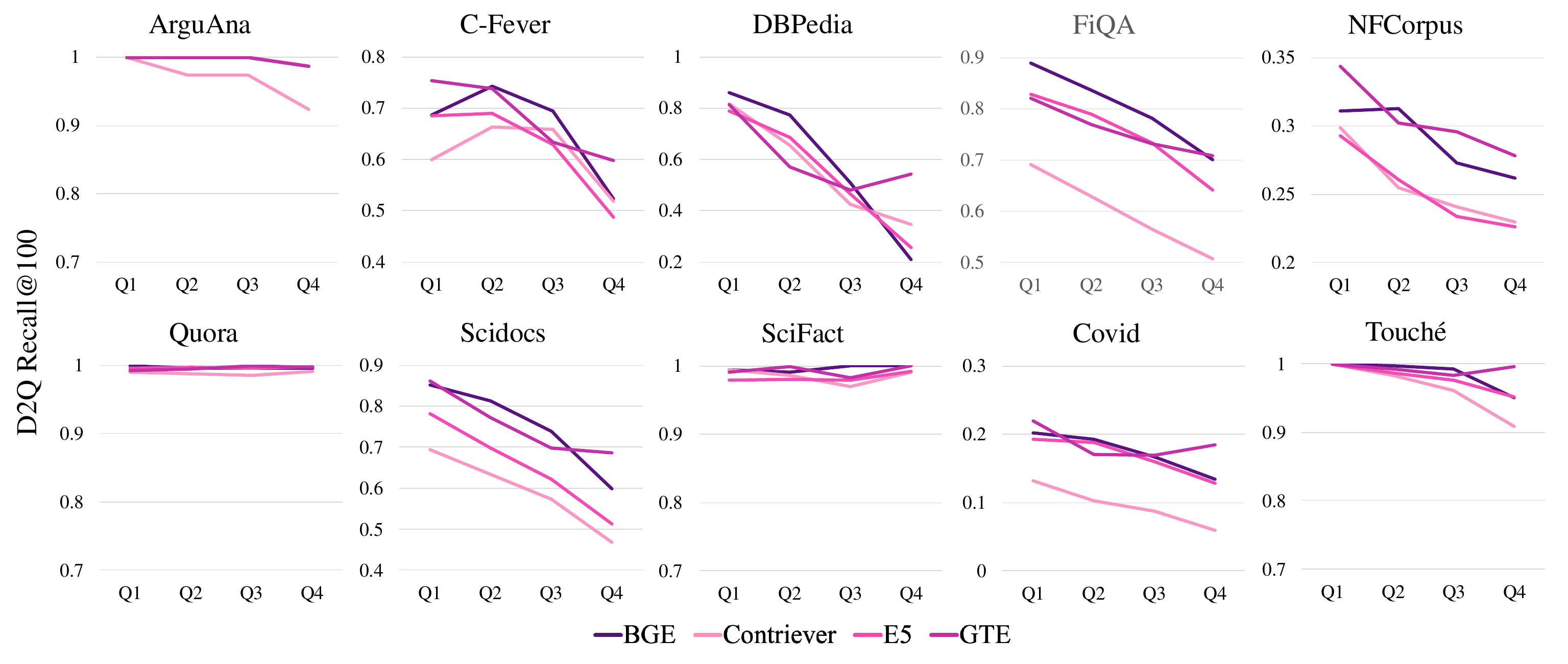}}
    \caption{Feasibility results of GradNormIR for several recent retrievers on the BEIR benchmark. The x-axis shows quartiles of GradNormIR, sorted in ascending order (Q1 to Q4), while the y-axis represents the d2q recall@100, averaged across documents within each quartile. The results show that GradNormIR can predict retrieval performance; lower GradNormIR values (Q1) generally lead to better retrieval outcomes across most datasets. As GradNormIR increases (Q4), the d2q recall decreases.
    }
    \label{fig:exp-d2q}
\end{figure*}

We aim to validate whether GradNormIR can identify the documents that are difficult for the the models to retrieve. To this end, we inspect if there is a consistent relationship between the computed gradient norm and the likelihood of a document successfully retrieved by its associated queries.

\textbf{Evaluation Metric.}
To evaluate the effectiveness, we measure the document-to-query (d2q) as the standard metric. In each dataset, annotations are provided in the form of $\{q_i, D_i\}_{i=1}^N$, where $q_i$ is a query and $D_i$ is the set of relevant documents. We reorganize these annotations as $\{d_i, Q_i\}_{i=1}^N$, where $Q_i$ represents the set of relevant queries for each document $d_i$.
For a document to be considered effectively retrievable, it should be retrieved for all its relevant queries. 

To quantify this, we define the d2q recall as follows: 
\begin{equation}
\text{recall}_\text{d2q} = \frac{\sum_{q_i \in Q_i} \mathbbm{I} \{d_i \in D^+(q_i)\}}{|Q_i|},    
\end{equation}
where $\mathbbm{I}$ is an indicator function and $D^+(q_i)$ represents the top-$k$ retrieved documents (with $k=100$). 

When the retriever model generalizes well for a document $d_i$, the d2q recall value will be high. Additionally, if the retriever generalizes effectively on $d_i$, the gradient norm associated with $d_i$ will be low, as the retriever does not need to make substantial updates based on the contrastive loss for $d_i$. Therefore, there should be an inverse relationship: higher the d2q recall values correspond to lower the gradient norms.

\textbf{Results.} Figure~\ref{fig:exp-d2q} illustrates the relationship between GradNormIR and d2q recall. We divide the data points into quartiles based on GradNormIR values, sorted in ascending order and labeled as Q1, Q2, Q3, and Q4. The x-axis represents these quartiles, while the y-axis shows the average d2q recall for each group. 

The results reveal a strong inverse correlation between GradNormIR and retrieval performance. As GradNormIR values increase from Q1 to Q4, d2q recall decreases. This indicates that higher GradNormIR values (Q4) are associated with documents that are more challenging for the retriever to retrieve consistently. Conversely, lower GradNormIR values (Q1) correspond to higher recall, indicating better retrieval performance. When d2q recall approaches 1, such as Quora and SciFact, this trend becomes less noticeable. This is likely because the datasets have been trained on; nearly all documents are well generalized and easily retrievable. 

\end{document}